\documentstyle[12pt]{article}
\begin{document}
\topmargin -0.2cm \oddsidemargin -0.2cm \evensidemargin -1cm
\textheight 22cm \textwidth 12cm

\title{New resonance-polariton Bose-quasiparticles enhances optical transmission into nanoholes in metal films.}
\author{Minasyan V.N. and Samoilov V.N.\\
Scientific Center of Applied Research, JINR,\\
Dubna, 141980, Russia}

\date{\today}

\maketitle

\begin{abstract}
We argue the existence of fundamental particles in nature, neutral
Light-Particles with spin 1, and rest mass $m=1.8\cdot 10^{-4} m_e$,
in addition to electrons, neutrons and protons. We call these
particles Light Bosons because they create the electromagnetic field
which represents Planck's gas of massless photons together with a
gas of Light Particles in the condensate. In this respect, the
condensed Light Particles, having no magnetic field, represent the
constant electric field. In this context, we predict a existence of
plasmon-polariton and resonance-polariton Bose-quasiparticles with effective masses
$m_l\approx 10^{-6} m_e$ and $m_r=0.5m_e$, which are induced by
interaction of the plasmon field and the resonance Fr$\ddot o$lich-
Schafroth charged bosons with electromagnetic wave in metal. Also,
we prove that the enhancement optical transmission into nanoholes in
metal films and Surface Enhanced Raman Spectroscopy are provided by
a new resonance-polariton Bose-quasiparticles but not model of
surface plasmon-polariton. In this letter, the quantization
Fresnel's equations is presented which confirms that Light Particles
in the condensate are concentrated near on the wall of grooves in
metallic grating and, in turn, represent as the constant electric
field which provides the launching of the surface Fr$\ddot o$lich-
Schafroth bosons on the surface metal holes.
\end{abstract}

\vspace{5mm}

{\bf 1. Introduction.}

\vspace{5mm}

In the letter  [1], we investigated the quantization scheme for the
local electromagnetic field which in turn concludes the existence of
light quasi-particles with spin $1$ and finite effective mass $m=
2.5\cdot 10^{-5} m_e$ (where $m_e$ is the mass of electron). We
showed that the Bose-gas consisting of the light quasi-particles in
homogenous medium can be described by the Bose- gas of two sort
polaritons. This picture is like to investigation of the superfluid
liquid $^4$He which was based on the London model [2] where
connection between the ideal Bose gas and superfluidity in liquid
$^4$He  was proposed. The ideal Bose gas undergoes a phase
transition at sufficiently low temperatures to a condition in which
the zero-momentum quantum state is occupied by a finite fraction of
the atoms. Later, Landau described the properties of superfluid
$^4$He in terms of collective excitations so-called phonons and
rotons [3]. The purely microscopic theory was first described by
Bogoliubov [4] within the model of weakly non-ideal Bose-gas, with
the inter-particle S- wave scattering. The Bogoliubov model for
superfluid helium have given a description of the gas consisting of
atoms helium via Bogoliubov's quasiparticles which reproduces
phonons at lower momenta (Landau prediction) as well as free atoms
at higher momenta (London's model of ideal gas). As we suggest the
picture of describing of superfluidity in liquid $^4$He is strongly
like to the problem of the model of electromagnetic field.
Consequently, there is arising the back problem which is connected
with understanding of the essence of electromagnetic wave because we
wish to know the origin of particles which may induce the Plank's
Bose-gas of massless photons.

For beginning, we shall examine the quantization scheme for local
electromagnetic field in the vacuum, as first presented by Planck in
his black body radiation studies. It is well known, that  classical
electrodynamics  leads to appearance of the so-called ultraviolet
catastrophe; to remove this problem, Planck modeled the
electromagnetic field as an ideal Bose gas of massless photons with
spin $1$. Later, Dirac [5] showed the Planck photon-gas could be
obtained through a quantization scheme for the local electromagnetic
field, by using of a model Bose-gas of local plane electromagnetic
waves.

It is well known, in quantum mechanics, a matter wave is determined
by wave-particle duality or de Broglie wave of matter [6] which was
confirmed by famous Davisson and Germer experiment, and also by
Compton effect [7] where the particle nature of light was
demonstrated. This reasoning allows us to present a model of
electromagnetic field as the non-ideal Bose-gas consisting of the
Bose-particles with spin $1$ and having non-zero rest mass. This
model is based on the application of the principles of the
wave-particle duality and the gauge invariance. As we show this
approach allows us to argue the existence of the fundamental neutral
Particles with spin $1$ and mass $m=1.8\cdot 10^{-4} m_e$ in nature,
which will be found by experiment. We call these particles as
Light-Particles because they create the gas of massless photons with
spin $1$.

There have been many studies of optical light transmission through
individual nanometer-sized holes in opaque metal films in recent
years [8-10]. These experiments showed highly unusual transmission
properties of metal films perforated with a periodic array of
subwavelength holes, because the electric field is highly localized
inside the grooves (around 300-1000 times larger than intensity of
incoming optical light). Here we analyze the absorption anomalies
for light in the visible to near-infrared range observed into
nanoholes in metal films. These absorption anomalies for optical
light as it seen as enhanced transmission of optical light in metal
films is provided by new resonance effect which is differ from model
surface plasmon polaritons (SPP)(collective electron density waves
propagating along the surface of the metal films) excited by light
incident on the hole array [11].

Hence, we mention about the Raman scattering which is a relatively
weak process. The number of photons Raman scattered is quite small.
However, as it is known the increase in intensity of the Raman
signal for adsorbates on surfaces occurs because of an enhancement
in the electric field provided by the surface. Surface Enhanced
Raman Spectroscopy, so called SERS, where results in the enhancement
of Raman scattering by molecules adsorbed on rough metal surfaces
[12,13]. In this context, we show when the incident light strikes
the surface of metal, localized new Bose-quasiparticles so called
resonance-polaritons are excited. The electric field enhancement is
greatest when the frequency of incoming photons is in resonance with
the Fr$\ddot o$lich- Schafroth's bosons. These resonance-polaritons
directed along to the surface where they are concentrated.

In this letter, we prove that the absorption anomalies for optical
light connected with presence new surface resonance-polaritons but
not by SPP as it was accepted. We also demonstrate that new
resonance-polariton Bose-quasiparticles in lamellar metallic
gratings lead to existence of Light Particles in the condensate,
when almost all Light Particles, incoming from air to interface on
metal, are concentrated near on the metal-air interface and fall the
condensate level. In turn, the Light Particles in the condensate on
the metal surface as constant electric field attribute to surface
Fr$\ddot o$lich- Schafroth's bosons for moving along the surface of
metal.

\vspace{5mm}

{\bf 11. Light-Particles with spin $1$ in vacuum.}

\vspace{5mm}

For beginning, we consider the Maxwell's equations for
electromagnetic field in vacuum [14]:
\begin{equation}
curl {\vec {H}} -\frac{1}{c}\frac{d {\vec{E}}}{d t}=0
\end{equation}

\begin{equation}
curl {\vec {E}} +\frac{1}{c}\frac{d {\vec{H}}}{d t}=0
\end{equation}

\begin{equation}
div {\vec {E}} =0
\end{equation}

\begin{equation}
div {\vec {H}} =0
\end{equation}

where $\vec {H}=\vec {H}(\vec {r},t)$ and $\vec {E}=\vec {E}(\vec
{r},t)$ are, respectively, the magnetic and electric field vectors
depending on space coordinate $\vec {r}$ and time $t$, and $c$ is
the velocity of light in vacuum. Obviously, $\varepsilon =1$ and
$\mu=1$ are, respectively, the dielectric and the magnetic
transmissivity of the vacuum.

The Hamiltonian of the radiation field $\hat{H}_R$ is:

\begin{equation}
\hat{H}_R =\frac{1}{8\pi }\int \biggl(E^2+ H^2\biggl) dV
\end{equation}

In order to solve a problem connected with a quantized
electromagnetic field in vacuum, we suggest that the electromagnetic
field in vacuum consists of $N$ Light-Particles with spin $1$ and
rest mass $m$ contained into box with volume $V$.
Due to application the principle of wave-particle duality,
we may suggest that these particles have the vectors of the electric
$\vec {E}_0=\vec {E}_0(\vec {r},t)$ and magnetic $\vec {H}_0=\vec
{H}_0(\vec {r},t)$ fields depending on space coordinate $\vec {r}$
and time $t$, and turn satisfy
to the Maxwell equations:

\begin{equation}
curl {\vec {H}_0} -\frac{1}{c}\frac{\partial {\vec{E}_0}}{\partial t}=0
\end{equation}

\begin{equation}
curl {\vec {E}_0} +\frac{1}{c}\frac{\partial {\vec{H}_0}}{\partial t}=0
\end{equation}

\begin{equation}
div {\vec {E}_0} =0
\end{equation}

\begin{equation}
div {\vec {H}_0} =0
\end{equation}

To find the relationship between the vectors  $\vec {E}$, $\vec {H}$
and $\vec {E}_0$, $\vec {H}_0$,  we introduce the following
expressions which in turn is provided the principle of the gauge
invariance:

\begin{equation}
\vec {E}=\alpha curl\vec {E}_0+\beta\cdot \vec {E}_0
\end{equation}
and
\begin{equation}
\vec {H}= \alpha curl\vec {H}_0 + \beta \vec{H}_0
\end{equation}
where $\alpha$ and $\beta$ are the unknown constants.

Thus, the electric $\vec {E}$ and magnetic $\vec {H}$ vectors of
initial electromagnetic field are determined by the secondary
electromagnetic field with electric $\vec {E}_0$ and
magnetic $\vec {H}_0$ vectors of the Light-Particles.
Obviously, $\vec {E}_0$ and $\vec {H}_0$ besides having relationship presented by (6) and (7), they
satisfy to the wave-equations:
\begin{equation}
\nabla^2 {\vec {E}_0}-\frac{1 }{c^2}\frac{\partial^2 \vec{E}_0}{\partial t^2}=0
\end{equation}

\begin{equation}
\nabla^2 {\vec {H}_0}-\frac{1}{c^2}\frac{\partial^2 \vec{H}_0}{\partial t^2}=0
\end{equation}

Since, we intend to present the quantized forms for electric and
magnetic operator-vectors $\vec {E}_0(\vec {r},t)$ and $\vec
{H}_0(\vec {r},t)$ of the Light-Particles which are propagated by
direction of wave normal $\vec {s}$, we postulate that the
directions of vector-operators of quantization electric and magnetic
fields of Light-Particles, and direction of propagation $\vec {s}$
are not changed, at introducing quantization scheme. Now, we express
the operator-vector $\vec {E}_0(\vec {r},t)$ electric field via the
second quantization vector wave functions of Light Boson:

\begin{equation}
\vec {E}_0(\vec{r}, t)= A\biggl(\vec{ \psi }(\vec{r}, t)
+\vec{\psi}^{+} (\vec{r}, t)\biggl)
\end{equation}

where $A$ is the norm coefficient; $\vec{\psi} (\vec{r}, t)$ and
$\vec{\psi}^{+} (\vec{r}, t)$ are, respectively, the second
quantization wave vector functions for one Light-Particle in point
of coordinate $\vec{r}$ and time $t$. These vector functions are
creation and annihilation operators of the electric field of one
Light-Particle in space coordinate-time, which is directed toward to
unit vector $\vec{e}$ perpendicularly to wave normal $\vec {s}$:
\begin{equation}
\vec{\psi} (\vec{r}, t)=
\frac{1}{\sqrt{V}}\sum_{\vec{k}}\vec{a}_{\vec{k},} e^{i(
k\vec{s}\vec{r} + \omega_{\vec{k}} t )}
\end{equation}

\begin{equation}
\vec{\psi}^{+} (\vec{r}, t)=
\frac{1}{\sqrt{V}}\sum_{\vec{k}}\vec{a}^{+}_{\vec{k}} e^{-i(
k\vec{s}\vec{r} + \omega_{\vec{k}}t )}
\end{equation}

and

\begin{equation}
\int \vec{\psi}^{+} (\vec{r}, t)\vec{\psi} (\vec{r}, t) dV=
\hat{N}_0+\sum_{\vec{k}\not=0}\vec{a}^{+}_{\vec{k}}
\vec{a}_{\vec{k}}=\hat{N}
\end{equation}

at

$$
\frac{1}{V} \int e^{i\vec{k}\cdot\vec{r}}dV=\delta_{\vec{k}}
$$

where  $\vec{a}^{+}_{\vec{k}}$ and $\vec{a}_{\vec{k}}$ are,
respectively, the Bose vector-operators of creation and annihilation
of quantized electric field of one Light-Particle in vacuum with
wave vector $\vec{k}$ which is directed along wave normal $\vec{s}$
or $\vec{k}=k\vec{s}$. These Bose vector-operators are directed to
the direction of the unit vector $\vec{e}$ which is perpendicular to
wave normal $\vec{s}$; $\hat{N}$ is the operator total number of
light bosons in vacuum; $\hat{N}_0=\hat{a}^{+}_0\hat{a}_0$  is the
total number of light particles in the condensate.

While investigating superfluid liquid, Bogoliubov [4] separated the
atoms of helium in the condensate from those atoms filling states
above the condensate. In an analogous manner, we may consider the
vector operators  $\hat{a}_0=\vec{e}\sqrt{N_0}$ and
$\hat{a}^{+}_0=\vec{e}\sqrt{N_0}$ as c-numbers  within the
approximation of a macroscopic number of Light- Particles in the
condensate $N_0\gg 1$. This assumptions lead to a broken
Bose-symmetry law for Light-Particles in the condensate. Taking the
given conclusion, we get

\begin{equation}
\vec {E}_0(\vec{r}, t)= \vec
{E}_{0,0}+\frac{A}{\sqrt{V}}\sum_{\vec{k}\not=0}\biggl( \vec {a}
_{\vec{k}}e^{i(k\vec{s}\vec{r} + kct)}  +\vec {a}^{+}_{\vec{k}}
e^{-i(k\vec{s}\vec{r} + kct)}\biggl)
\end{equation}

where $\vec {E}_{0,0}=2A\vec{e}\sqrt{\frac{N_0}{V}}$.

Application (18) into (7), by taking into account (13), leads to

\begin{equation}
\vec {H}_0(\vec{r}, t)=\vec
{H}_{0,0}-\frac{A}{\sqrt{V}}\sum_{\vec{k}\not=0}\biggl(
\vec{s}\times\vec {a} _{\vec{k}} e^{kct} - \vec{s}\times\vec
{a}^{+}_{-\vec{k}} e^{-i kct}\biggl)e^{i\vec{k}\vec{r}}
\end{equation}

where
$$
\vec {H}_{0,0}=-A\vec{s}\times
\vec{e}\biggl(\sqrt{\frac{N_0}{V}}-\sqrt{\frac{N_0}{V}}\biggl)=0
$$

In fact, the Light-Particles in the condensate reproduce only the
constant electric field $\vec
{E}_{0,0}=2A\vec{e}\sqrt{\frac{N_0}{V}}$ which is directed along
unit vector $\vec{e}$.

In this context, there is an important condition for transverse
electromagnetic field $\vec {E}_0\cdot \vec {H}_0=0$ which is easy
to prove by using (18) and (19), and equality
$\vec{a}(\vec{b}\times\vec {c})=\vec{c}(\vec{a}\times\vec
{b})=\vec{b}(\vec{c}\times\vec {a})$ and $\vec{a}\times\vec
{b}=-\vec{b}\times\vec {a}$.

Further calculation claims to find the operators $\frac{\partial
{\vec{E}_0}(\vec{r}, t)}{\partial t}$ and $\frac{\partial
{\vec{H}_0}(\vec{r}, t)}{\partial t}$ which by prescription of Dirac
[5], at current time $t=0$, they take the forms:

\begin{equation}
\frac{\partial {\vec{E}_0}}{\partial
t}=\frac{icA}{\sqrt{V}}\sum_{\vec{k}}k\biggl(\vec{a}_{\vec{k}}-\vec{a}^{+}_{-\vec{k}}\biggl)e^{i\vec{k}\vec{r}}
\end{equation}

\begin{equation}
\frac{\partial {\vec{H}_0}}{\partial
t}=\frac{icA}{\sqrt{V}}\sum_{\vec{k}}k\vec{s}\times\biggl(\vec{a}_{\vec{k}}+\vec{a}^{+}_{-\vec{k}}
\biggl)e^{i\vec{k}\vec{r}}
\end{equation}

With these new terms $\vec {E}_0$ and $\vec {H}_0$ by using of (6)
and (7) into (10) and (11), the radiation Hamiltonian $\hat{H}_R$ in
(5) takes the form:

\begin{equation}
\begin{array}{ll}
\hat{H}_R =\frac{1}{8\pi}\int \biggl( E^2+H^2\biggl) dV =\\[+12pt]
\displaystyle
=\;\frac{1}{8\pi}\int \biggl [\biggl(- \frac{\alpha}{c}\frac{\partial {\vec{H}_0}}{\partial t}+\beta \vec {E}_0\biggl)^2+\\[+12pt]
\displaystyle +\;\biggl(\frac{\alpha }{c}\frac{\partial {\vec{E}_0}}{\partial t}+
\beta \vec {H}_0\biggl)^2\biggl] dV
\end{array}
\end{equation}

Now, we introduce an expression
$$
(\vec{a}\times\vec{b})(\vec{c}\times\vec{d})=(\vec{a}\cdot\vec{c})(\vec{b}\cdot\vec{d})-(\vec{a}\cdot\vec{d})(\vec{b}\cdot\vec{c})
$$
and a condition for transverse electromagnetic field
$\vec{s}\cdot\vec{a}_{\vec{k}}=0$ and
$\vec{s}\cdot\vec{a}^{+}_{\vec{k}}=0$, as result of application (18)
into (8), which helps us to obtain the reduced form of $\hat{H}_R$
in (22). Hence, it is a necessary to indicate that expansion of the
vectors $\vec {E}_0(\vec{r}, t)$ and $\vec {H}_0(\vec{r}, t)$ by
wave vector $\vec{k}$ is correct, if they are given by different
wave vectors $\vec{k}$ and $\vec{k}^{'}$, respectively. However, the
finally result requests an accepting $\vec{k}=\vec{k}^{'}$. Thus

\begin{equation}
\hat{H}_R=\hat{H}_a+\hat{H}_b
\end{equation}

where the operators $\hat{H}_a $ and $\hat{H}_b$ are:

\begin{equation}
\hat{H}_a =\frac{1}{8\pi}\int \biggl[\frac{\alpha^2}{c^2}
\biggl(\frac{\partial {\vec{H}_0}}{\partial t}\biggl)^2- \frac{2\alpha\beta}{c}
\frac{\partial {\vec{H}_0}}{\partial t} \vec {E}_0+\beta^2 \vec {E}^2_0\biggl)^2\biggl]dV
\end{equation}
and
\begin{equation}
\hat{H}_b =\frac{1}{8\pi}\int \biggl[\frac{\alpha^2}{c^2}
\biggl(\frac{\partial {\vec{E}_0}}{\partial t}\biggl)^2+\frac{2\alpha \beta }{c}
\frac{\partial {\vec{E}_0}}{\partial t}\vec{H}_0+\beta^2 \vec {H}^2_0\biggl)^2\biggl]dV
\end{equation}
with the terms involving  above equations which are

\begin{equation}
\frac{1}{8\pi}\int \frac{\alpha^2}{c^2} \biggl(\frac{\partial
{\vec{H}_0}}{\partial t}\biggl)^2 dV=
\sum_{\vec{k}^{'}}\frac{\alpha^2 A^2
{k^{'}}^2}{8\pi}\biggl(\vec{a}_{\vec{k}^{'}}+\vec{a}^{+}_{-\vec{k}^{'}}\biggl)\biggl(\vec{a}_{-\vec{k}^{'}}+
\vec{a}^{+}_{\vec{k}^{'}}\biggl)
\end{equation}

\begin{equation}
\frac{1}{8\pi}\int \frac{\alpha^2}{c^2}
\biggl(\frac{\partial {\vec{E}_0}}{\partial t}\biggl)^2 dV=-
\sum_{\vec{k}}\frac{\alpha^2 A^2 k^2}{8\pi}\biggl(\vec{a}_{\vec{k}}-\vec{a}^{+}_{-\vec{k}}\biggl)\biggl(\vec{a}_{-\vec{k}}-\vec{a}^{+}_{\vec{k}}\biggl)
\end{equation}

\begin{equation}
\frac{1}{8\pi}\int \beta^2 \vec {E}^2_0 dV=
\sum_{\vec{k}}\frac{\beta^2}{8\pi}\biggl(\vec{a}_{\vec{k}}+\vec{a}^{+}_{-\vec{k}}\biggl)
\biggl(\vec{a}_{-\vec{k}}+\vec{a}^{+}_{\vec{k}}\biggl)
\end{equation}

\begin{equation}
\frac{1}{8\pi}\int \beta^2 \vec {H}^2_0 dV=
\sum_{\vec{k}^{'}}\frac{\beta^2}{8\pi}\biggl(\vec{a}_{\vec{k}^{'}}-\vec{a}^{+}_{-\vec{k}^{'}}\biggl)
\biggl(\vec{a}_{-\vec{k}^{'}}-\vec{a}^{+}_{\vec{k}^{'}}\biggl)
\end{equation}

\begin{equation}
\frac{2\alpha\beta}{c}\frac{d {\vec{H}_0}}{d t} \vec {E}_0=\frac{2\alpha \beta }{c}\frac{d {\vec{E}_0}}{d t}\vec{H}_0=0
\end{equation}

If we suggest that $\alpha=\frac{\hbar \sqrt{\pi}}{A\sqrt{2m}}$,
then, the Eq.(23) may rewrite down by approximation of a macroscopic
number of Light- Particles in the condensate $N_0\gg 1$, as

\begin{equation}
\begin{array}{ll}
\hat{H}_R =\frac{\beta^2 A^2 N_0}{2\pi}+\sum_{\vec{k}\not=0}\biggl
[\biggl (\frac{\hbar^2 k^2 }{2m }+ \frac{\beta^2 A^2}{4\pi}\biggl )
\vec{a}^{+}_{\vec{k}}\vec{a}_{\vec{k}}-\\[+12pt]
\displaystyle -\;\frac{1}{2}\biggl (\frac{\hbar^2 k^2 }{2m }-
\frac{\beta^2 A^2}{4\pi}\biggl )\biggl (\vec{a}^{+}_{\vec{k}}
\vec{a}^{+}_{-\vec{k}}+
\vec{a}_{-\vec{k}}\vec{a}_{\vec{k}}\biggl)\biggl ]+\\[+12pt]
\displaystyle +\;\sum_{\vec{k}^{'}\not=0}\biggl[\biggl(\frac{\hbar^2
{k^{'}}^2 }{2m }+ \frac{\beta^2 A^2}{4\pi}\biggl )
\vec{a}^{+}_{\vec{k}^{'}}\vec{a}_{\vec{k}^{'}}+\\[+12pt]
\displaystyle +\;\frac{1}{2}\biggl (\frac{\hbar^2 {k^{'}}^2 }{2m }-
\frac{\beta^2 A^2}{4\pi}\biggl ) \biggl (\vec{a}^{+}_{\vec{k}^{'}}
\vec{a}^{+}_{-\vec{k}^{'}}+
\vec{a}_{-\vec{k}^{'}}\vec{a}_{\vec{k}}^{'}\biggl)\biggl ]
\end{array}
\end{equation}

We now apply new transformation for vector-operator [1], which is similar to Bogoliubov's one [4], for evaluation the
energy levels of the operator $\hat{H}_R$ within diagonal form:
\begin{equation}
\vec{a}_{\vec{k}}=\frac{\vec{i}_{\vec{k}} +
L_{\vec{k}}\vec{i}^{+}_{-\vec{k}}} {\sqrt{1-L^2_{\vec{k}}}}
\end{equation}

where $L_{\vec{k}}$ is the real symmetrical
functions of a wave vector $\vec{k}$.

The operator Hamiltonian $\hat{H}_R$ by using of a canonical
transformation takes a following form:
\begin{equation}
\hat{H}_R= \frac{\beta^2 A^2 N_0}{2\pi}+\sum_{
\vec{k}}\chi_{\vec{k}}\vec{i}^{+}_{\vec{k}} \vec{i}_{\vec{k}}+\sum_{
\vec{k}^{'}}\chi_{\vec{k}^{'}}\vec{i}^{+}_{\vec{k}^{'}}
\vec{i}_{\vec{k}^{'}}
\end{equation}

Hence, we infer that the Bose-operators $\vec{i}^{+}_{\vec{k}}$,
$\vec{i}^{+}_{\vec{k}^{'}}$ and $\vec{i}_{\vec{k}}$,
$\vec{i}_{\vec{k}^{'}}$ are, respectively, the vector "creation" and
"annihilation" operators of massless photons with energies
$\chi_{\vec{k}}$ and $\chi_{\vec{k}^{'}}$:

\begin{equation}
\begin{array}{ll}
\chi_{\vec{k}}=\sqrt{\biggl (\frac{\hbar^2k^2}{2m}+\frac{\beta^2
A^2}{4\pi}\biggl )^2-\biggl (\frac{\hbar^2k^2}{2m}-\frac{\beta^2
A^2}{4\pi}\biggl)^2}=\\[+12pt]
\displaystyle =\;\frac{\hbar k \beta A}{\sqrt{2m\pi}}=\hbar k c
\end{array}
\end{equation}

and

\begin{equation}
\begin{array}{ll}
\chi_{\vec{k}^{'}}=\sqrt{\biggl (\frac{\hbar^2k^2}{2m}+\frac{\beta^2
A^2}{4\pi}\biggl )^2-\biggl (\frac{\hbar^2k^2}{2m}-\frac{\beta^2
A^2}{4\pi}\biggl)^2}=\\[+12pt]
\displaystyle =\;\frac{\hbar k^{'} \beta A}{\sqrt{2m\pi}}=\hbar k^{'} c
\end{array}
\end{equation}

where the constant $\beta=\frac{c\sqrt{2m\pi}}{A}$ is defined.

Taking $\vec{k}=\vec{k}^{'}$, we obtain
$\chi_{\vec{k}}=\chi_{\vec{k}^{'}}$, and

\begin{equation}
\hat{H}_R=mc^2 N_0+2\sum_{
\vec{k}\not=0}\chi_{\vec{k}}\vec{i}^{+}_{\vec{k}} \vec{i}_{\vec{k}}
\end{equation}

where $mc^2 N_0$ is new term, in regard to Plank's formulae, which
determines the energy of Light-Particles in the condensate. This
reasoning implies that the Light-Particles in the condensate
represent as the constant electric field without magnetic one.

Hence, we note that the operator $\hat{H}_R$ in (31) may present as
\begin{equation}
\hat{H}_R =\hat{H}_e + \hat{H}_h
\end{equation}

where
\begin{equation}
\hat{H}_e=\sum_{\vec{k}}\biggl(\frac{\hbar^2 k^2 }{2m }+
\frac{mc^2}{2}\biggl)
\vec{a}^{+}_{\vec{k}}\vec{a}_{\vec{k}}+\frac{1}{2}\sum_{\vec{k}}U_{\vec{k}}
\biggl (\vec{a}^{+}_{\vec{k}} \vec{a}^{+}_{-\vec{k}}+
\vec{a}_{-\vec{k}}\vec{a}_{\vec{k}}\biggl)
\end{equation}
and

\begin{equation}
\hat{H}_h=\sum_{\vec{k}}\biggl(\frac{\hbar^2 {k^{'}}^2 }{2m }+
\frac{mc^2}{2}\biggl)
\vec{a}^{+}_{\vec{k}^{'}}\vec{a}_{\vec{k}^{'}}+\frac{1}{2}\sum_{\vec{k}^{'}}U_{\vec{k}^{'}}
\biggl (\vec{a}^{+}_{\vec{k}^{'}} \vec{a}^{+}_{-\vec{k}^{'}}+
\vec{a}_{-\vec{k}^{'}}\vec{a}_{\vec{k}^{'}}\biggl)
\end{equation}

where

\begin{equation}
U_{\vec{k}}=-\frac{\hbar^2k^2} {2m}+\frac{mc^2}{2}
\end{equation}

\begin{equation}
U_{\vec{k}^{'}}=\frac{\hbar^2{k^{'}}^2}{2m}-\frac{mc^2}{2}
\end{equation}

are the potential interaction between the Light-Particles in wave
vector space.

We argue that there is a condition for wave numbers of
Light-Particles $k< k_0$ which provides the property of hard
particles for Light Particle. This condition requests that the
potentials $U_{\vec{k}}$ and $U_{\vec{k}^{'}}$ interaction between
particles were a repulsive, which must occur namely for particles
which having the form of hard sphere form (recall S-wave repulsive
pseudopotential interaction between atoms in the superfluid liquid
$^4$He in the model hard spheres [15]). However, when

\begin{equation}
U_{\vec{k}}=-\frac{\hbar^2k^2} {2m}+ \frac{mc^2}{2} >0
\end{equation}
then

\begin{equation}
U_{\vec{k}^{'}}=\frac{\hbar^2{k^{'}}^2} {2m}-\frac{mc^2}{2}<0
\end{equation}

These equations determine the boundary wave number

\begin{equation}
k_0=\frac{m c}{\hbar}
\end{equation}

for Light-Particles which separates the particles by repulsive and
attractive potentials interaction because for Light Particles, with
wave vectors belonging to $0\leq k< k_0$, are separated by the
repulsive potential $U_{\vec{k}}$ and by attractive
$U_{\vec{k}^{'}}$ which, respectively, determine  but for Light
Particles with wave number determines the Light Particles with
$U_{\vec{k}}>0$ as Particles with form of hard sphere, and the Light
Particles with $U_{\vec{k}^{'}}<0$ as Quasiparticles because they
may not satisfy to condition of model hard sphere. The separation of
the Light Particles on Particles and Quasiparticles which
respectively correspond to Particles of electric field and
Quasiparticles to magnetic field is confirmed by reasoning that
electric field consists of the Light Bosons in the condensate (18)
but magnetic field consists of  Light Quasiparticles which cannot
set into the condensate (19).

Thus, in statistical equilibrium or thermodynamic limit, the average
energy of system is presented as

\begin{equation}
\overline{\hat{H}}_R=mc^2 N_0+2\sum_{ 0\leq k<
k_0}\chi_{\vec{k}}\vec{i}^{+}_{\vec{k}} \vec{i}_{\vec{k}}
\end{equation}

where $\overline{\vec{i}^{+}_{\vec{k}}\vec{i}_{\vec{k}}}$ is the
average number of photons  with the wave vector $\vec{k}$ at
temperature $T$:

\begin{equation}
\overline{\vec{i}^{+}_{\vec{k}}\vec{i}_{\vec{k}}}=
\frac{1}{e^{\frac{\hbar k c}{kT}}-1}
\end{equation}

Now, we prove that the existence of Light Particles satisfies to the
relativistic theory of Einstein. Considering Light-Particle as de
Broglie wave, we may express the boundary wave number $k_0$ by
following form:

\begin{equation}
k_0=\frac{m
v_{max}}{\hbar\sqrt{1-\frac{v^2_{max}}{c^2}}}
\end{equation}

where $v_{max}$ is the maximal value speed of the Light-Particle.
Substituting quantity boundary wave number $k_0$ from (45) into
(47), we obtain the following equation
\begin{equation}
\frac{m c}{\hbar}=\frac{m
v_{max}}{\hbar\sqrt{1-\frac{v^2_{max}}{c^2}}}
\end{equation}

which determines a quantity $v_{max}=\frac{c}{\sqrt{2}}<c$. In this
context, there is a Mistaken into the paper [16] where it was stated
that Relativistic theory, proposed by Einstein, is incorrect. Just,
we showed that the definition of object, as Light-Particles, within
model hard sphere, leads to Right Sound of the theory of Einstein,
because the maximal value of speed of the Light-Particle is less
then the velocity of electromagnetic wave in vacuum
$v_{max}=\frac{c}{\sqrt{2}}<c$.

Now, it is a necessary to find the mass of Light Boson which it will
be calculated in section IX as experimental result.

\vspace{5mm}

{\bf 111. The gauge invariance.}

\vspace{5mm}

Thus,we may recall that solutions of (1)-(4) and (6)-(9) have the
forms:

\begin{equation}
\vec {E}=-\frac{\partial\vec{A}}{c\partial t}-grad\phi
\end{equation}

\begin{equation}
\vec {H}=curl\vec{A}
\end{equation}
and

\begin{equation}
\vec {E}_0=-\frac{\partial\vec{A}_0}{c\partial t}-grad\phi_0
\end{equation}

\begin{equation}
\vec {H}_0=curl\vec{A}_0
\end{equation}

where $\vec{A}$ and $\phi$ are, respectively,the  vector  and scalar
potentials of initial electromagnetic field; $\vec{A}_0$ and
$\phi_0$ are, respectively,the vector and scalar potentials of
secondary electromagnetic field or electromagnetic field of Light
Bosons.

On other hand, there are presence two conditions of transversely
polarized excitations:

\begin{equation}
div \vec {A}=0
\end{equation}

and

\begin{equation}
div \vec {A}_0=0
\end{equation}

In empty space, far from
any electric charges, we usually assume that scalar potentials of
initial and secondary electromagnetic fields are zero $\phi=0$ and $\phi_0=0$. Then,

\begin{equation}
\vec {E}=-\frac{\partial \vec{A}}{c \partial t}
\end{equation}
and

\begin{equation}
\vec {E}_0=-\frac{\partial \vec{A}_0}{c\partial t}
\end{equation}

Therefore, inserting the vectors $\vec {H}$, $\vec {E}$ from
(50),(55) and $\vec {H}_0$, $\vec {E}_0$ from (52),(56) into (10)
and (11), we obtain
\begin{equation}
\vec {A}=\alpha curl\vec {A}_0 +\beta\vec {A}_0
\end{equation}

By using of the operation $div$ to the both sides of Eq.(57), and
accepting the Coulomb gauge condition for initial electromagnetic
field $div \vec {A}=0$ in (53), then we get the result $div \vec
{A}_0=0$ presented in (54) automatically, because $div(curl\vec
{A}_0 )=0$. Thus, the existence of Light Bosons is not contradicting
to the quantum electrodynamics because the Coulomb gauge condition
for secondary electromagnetic field is fulfilled.

\vspace{5mm}

{\bf 1V. Light-Particles with spin $1$ in homogenous medium.}

\vspace{5mm}

For beginning, we consider the Maxwell's equations for
electromagnetic field in homogenous medium which are presented by
the forms:

\begin{equation}
curl {\vec {H}} -\frac{1}{c}\frac{\partial {\vec{D}}}{\partial t}=0
\end{equation}

\begin{equation}
curl {\vec {E}} +\frac{1}{c}\frac{\partial {\vec{B}}}{\partial t}=0
\end{equation}

\begin{equation}
div {\vec {D}} =0
\end{equation}

\begin{equation}
div {\vec {B}} =0
\end{equation}

where $\vec {B}=\vec {B}(\vec {r},t)$ and $\vec {D}=\vec {D}(\vec
{r},t)$ are, respectively, the local magnetic and electric induction
depending on space coordinate $\vec {r}$ and time $t$; $\vec
{H}=\vec {H}(\vec {r},t)$ and $\vec {E}=\vec {E}(\vec {r},t)$ are,
respectively, the magnetic and electric field vectors, and $c$ is
the velocity of light in vacuum. The further equations are
\begin{equation}
\vec {D} =\varepsilon \vec {E}
\end{equation}

\begin{equation}
\vec {B} =\mu \vec {H}
\end{equation}

where $\varepsilon=\varepsilon(\vec {r}-\vec {r}^{'};t-t^{'})$ and
$\mu=1$ are, respectively, the dielectric and the magnetic
susceptibilities of the homogenous medium. Hence, we note that
dielectric respond $\varepsilon$ of the
homogenous medium is only a function of the separations $\vec
{r}-\vec {r}^{'}$ and $t-t^{'}$ from the external field at $\vec
{r}^{'}$ and $t^{'}$.

In analogy, manner, as it was made in above, the electric $\vec
{E}$, electric induction $\vec {D}$ and magnetic $\vec {H}$ vectors
of initial electromagnetic field are determined by the secondary
electromagnetic field with electric $\vec {E}_0$, electric induction
$\vec {D}_0$ and magnetic $\vec {H}_0$ vectors of the
Light-Particles in homogenous medium by application of the
principles of wave-particle duality and gauge invariance, which in
turn lead to relationships:

\begin{equation}
\vec {E}=\alpha curl\vec {E}_0+\beta\cdot \vec {E}_0
\end{equation}

\begin{equation}
\vec {D}=\alpha curl\vec {D}_0+\beta\cdot \vec {D}_0
\end{equation}
and
\begin{equation}
\vec {H}= \alpha curl\vec {H}_0 + \beta \vec{H}_0
\end{equation}

where $\alpha=\frac{\hbar \sqrt{\pi}}{A\sqrt{2m}}$ and $\beta=
\frac{2c\sqrt{2m\pi}}{A}$.

Thus, the Maxwell's equations for secondary electromagnetic field
are presented as

\begin{equation}
curl {\vec {H}_0} -\frac{1}{c}\frac{\partial {\vec{D}_0}}{\partial
t}=0
\end{equation}

\begin{equation}
curl {\vec {E}_0} +\frac{1}{c}\frac{\partial {\vec{H}_0}}{\partial t}=0
\end{equation}

\begin{equation}
div {\vec {D}_0} =0
\end{equation}

\begin{equation}
div {\vec {H}_0} =0
\end{equation}

with
\begin{equation}
\vec {D}_0 =\varepsilon \vec {E}_0
\end{equation}

\begin{equation}
\vec {B}_0 =\vec {H}_0
\end{equation}

By prescription of Kubo in [17] for electrical conductivity in
solids. In analogy manner, we treat the dielectric response
$\varepsilon(\vec {r}-\vec {r}^{'},t-t^{'})$ for homogenous medium.
In this respect, we determine the equations for electric induction
$\vec {D}_0(\vec {r},t)$ of plane electromagnetic wave, propagating
in fixed direction of unit vector $\vec {s}$ in homogeneous medium.
In this case, the electric induction $\vec {D}_0(t)$ may depend on
$\vec {E}_0(t^{'})$ in precess moment of time but not next one:

\begin{equation}
\vec {D}_0(\vec {r},t) =\int d^3
r^{'}\int^{t}_{-\infty}dt^{'}\varepsilon(\vec {r}-\vec {r}^{'};
t-t^{'})\vec {E}_0(\vec {r}^{'},t^{'})
\end{equation}

By application of the Fourier transformations:

\begin{equation}
\vec {D}_0(\vec {r},t)=\frac{1}{2\pi}
\int^{+\infty}_{0}d\omega_{\vec{k}} e^{-i \omega_{\vec{k}} t}\vec
{D}_0(\vec {r},\omega_{\vec{k}})
\end{equation}

\begin{equation}
\vec {D}_0(\vec {r},\omega_{\vec{k}})= \int dt e^{-i
\omega_{\vec{k}} t}\vec {D}_0(\vec {r},t)
\end{equation}
we obtain

\begin{equation}
\vec {D}_0(\vec {r},\omega)= \int dr^{'}\varepsilon(\vec {r}-\vec
{r}^{'};\omega_{\vec{k}})\vec {E}_0(\vec {r}^{'},\omega_{\vec{k}})
\end{equation}

where

\begin{equation}
\varepsilon(\vec {r}-\vec
{r}^{'};\omega_{\vec{k}})=\int^{\infty}_{0}dt\varepsilon(\vec
{r}-\vec {r}^{'}; t)e^{i \omega_{\vec{k}} t}
\end{equation}

Then, making the Fourier transformation by coordinate space, we
obtain in space wave number-frequency:

\begin{equation}
\vec {D}_0(\vec {k},\omega_{\vec{k}}) =\varepsilon(\vec {k},
\omega_{\vec{k}})\vec {E}_0(\vec {k},\omega_{\vec{k}})
\end{equation}
where

\begin{equation}
\vec {D}_0(\vec {k},\omega_{\vec{k}}) =\int
dre^{-i\vec{k}\vec{r}}\vec {D}_0(\vec {r},\omega_{\vec{k}})
\end{equation}

Thus, we have an important result
\begin{equation}
\vec {D}_0(\vec {r},t) =\sum_{\vec{k}}\varepsilon(\vec {k},
\omega_{\vec{k}})\vec {E}_0(\vec {k},\omega_{\vec{k}}) e^{i(
\vec{k}\vec{r} + \omega_{\vec{k}} t)}
\end{equation}

In this context, we postulate that the electric $\vec {E}_0(\vec
{r},t)$ operator-vectors of the Light-Particles is expressed via the
second quantization vector wave functions of Light Boson which
determines a creation and an annihilation the Light-Particle in
space coordinate-time by following way:

\begin{equation}
\vec {E}_0(\vec{r}, t)= A\biggl( \phi (\vec{r}, t) +\phi^{+}
(\vec{r}, t)\biggl)
\end{equation}

where $A$ is the scalar amplitude of electric field; $\vec{\phi}
(\vec{r}, t)$ and $\vec{\phi}^{+} (\vec{r}, t)$ are, respectively,
the second quantization wave vector functions for one Light-Particle
in homogenous medium in point of coordinate $\vec{r}$ and  current
time $t$:

\begin{equation}
\vec{\phi} (\vec{r}, t)=
\frac{1}{\sqrt{V}}\sum_{\vec{k}}\vec{b}_{\vec{k}} e^{i(
\vec{k}\vec{r} + \omega_{\vec{k}} t)}
\end{equation}

\begin{equation}
\vec{\phi}^{+} (\vec{r}, t)=
\frac{1}{\sqrt{V}}\sum_{\vec{k}}\vec{b}^{+}_{\vec{k}} e^{-i(
\vec{k}\vec{r}+ \omega_{\vec{k}} t)}
\end{equation}

and
\begin{equation}
\int\vec{\phi}^{+} (\vec{r}, t)\vec{\phi} (\vec{r}, t) dV =
\hat{N}_{0,h}+\sum_{\vec{k}\not=0}\vec{b}^{+}_{\vec{k}}
\vec{b}_{\vec{k}}=\hat{N}_h
\end{equation}

where  $\vec{b}^{+}_{\vec{k}}$ and $\vec{b}_{\vec{k}}$ are,
respectively, the Bose vector-operators of creation and annihilation
of the electric field for free one Light-Particles with spin $1$ in
homogeneous medium, described by a vector $\vec{k}$ and frequency
$\omega_{\vec{k}}$. The Bose vector-operators are directed by the
electric field along unit vector $\vec{e}$ which is perpendicular to
vector $\vec{k}$; $\hat{N}_h$ is the operator total number of Light
bosons in homogenous medium;
$\hat{N}_{0}=\hat{b}^{+}_{0}\hat{b}_{0}$ is the total number of
Light Particles in homogenous medium which fill the condensate level
$\vec{k}=0$.

In analogy manner, as it was presented in above, we may consider the
vector operators  $\hat{b}_0=\vec{e}\sqrt{N_0}$ and
$\hat{b}^{+}_0=\vec{e}\sqrt{N_{0,h}}$ as c-numbers  within the
approximation of a macroscopic number of Light-Particles in
homogeneous medium $N_{0,h}\gg 1$ which fill the condensate level.
This assumptions lead to a broken Bose-symmetry law for Light-
Particles in the condensate. Taking the given conclusion, we get

\begin{equation}
\vec {E}_0(\vec{r}, t)= \vec
{E}_{0,0}+\frac{A}{\sqrt{V}}\sum_{\vec{k}\not=0}\biggl( \vec {b}
_{\vec{k}}e^{i(\vec{k}\vec{r} + \omega_{\vec{k}} t)}  +\vec
{b}^{+}_{\vec{k}} e^{-i(\vec{k}\vec{r} + \omega_{\vec{k}} t)}\biggl)
\end{equation}

Then, by application (67), we obtain
\begin{equation}
\vec {D}_0= \vec
{D}_{0,0}+\frac{A}{\sqrt{V}}\sum_{\vec{k}\not=0}\varepsilon(\vec
{k},\omega_{\vec{k}})\biggl( \vec {b} _{\vec{k}}e^{i(\vec{k}\vec{r}
+ \omega_{\vec{k}} t)}  +\vec {b}^{+}_{\vec{k}} e^{-i(\vec{k}\vec{r}
+ \omega_{\vec{k}} t)}\biggl)
\end{equation}
where
$$
\vec {D}_{0,0}=2A\vec{e}\varepsilon(\vec {k}=0,\omega_{\vec
{k}=0})\sqrt{\frac{N_{0,h}}{V}}
$$

Obviously, the Eqs.(67)-(72) lead to the wave equation:

\begin{equation}
\nabla^2 {\vec {E}_0(\vec {r},t)}-\frac{1 }{c^2}\frac{\partial^2
\vec{D}_0(\vec {r},t)}{\partial t^2}=0
\end{equation}

where substituting $\vec {E}_0(\vec {r},t)$ and $\vec {D}_0(\vec
{r},t)$ from (85) and (86), we may get to the important equation for
Light Particles in homogenous medium:

\begin{equation}
\sqrt{\varepsilon(\vec {k},\omega_{\vec{k}})}
=\frac{kc}{\omega_{\vec{k}}}
\end{equation}

Obviously, at $\varepsilon(\vec {k},\omega_{\vec{k}})=1$, we deal
with electromagnetic field with frequency of Light Particle
$\omega_{\vec{k}}=kc$ presented into vacuum.

Thus,
\begin{equation}
\vec {H}_0=\vec
{H}_{0,0}-\frac{A}{\sqrt{V}}\sum_{\vec{k}\not=0}\sqrt{\varepsilon(\vec
{k},\omega_{\vec{k}})}\vec{s}\times\biggl(\vec {b} _{\vec{k}}
e^{i\omega_{\vec{k}} t} - \vec {b}^{+}_{-\vec{k}} e^{-i
\omega_{\vec{k}} t}\biggl)e^{i\vec{k}\vec{r}}
\end{equation}

where
$$
\vec {H}_{0,0}=-A\sqrt{\varepsilon(\vec
{k}=0,\omega_{\vec{k}=0})}\vec{s}\times
\vec{e}\biggl(\sqrt{\frac{N_{0,h}}{V}}-\sqrt{\frac{N_{0,h}}{V}}\biggl)=0
$$

Thus, the condensed Light Particles in homogeneous medium reproduce
the constant electric field $\vec
{E}_{0,0}=2A\vec{e}\sqrt{\frac{N_{0,h}}{V}}$ in direction $\vec{e}$
but no having a magnetic field.

Now, we use of a prescription of Dirac [5], at current time $t=0$,
the operators $\frac{\partial {\vec{D}_0}}{\partial t}$ and
$\frac{\partial {\vec{H}_0}}{\partial t}$ take the forms:

\begin{equation}
\frac{\partial {\vec{D}_0}}{\partial
t}=\frac{iAc}{\sqrt{V}}\sum_{\vec{k}}k \sqrt{\varepsilon(\vec
{k},\omega_{\vec{k}})}\biggl(\vec{b}_{\vec{k}}-\vec{b}^{+}_{-\vec{k}}\biggl)e^{i\vec{k}\vec{r}}
\end{equation}

and
\begin{equation}
\frac{\partial {\vec{H}_0}}{\partial
t}=\frac{iAc}{\sqrt{V}}\sum_{\vec{k}}\sqrt{\varepsilon(\vec
{k},\omega_{\vec{k}})}\vec{k}\times\biggl(\vec{b}_{\vec{k}}+\vec{b}^{+}_{-\vec{k}}
\biggl)e^{i\vec{k}\vec{r}}
\end{equation}

The Hamiltonian of the radiation field $\hat{H}_R$ in homogeneous
medium is very complex, as yet not fully solved. This is problem
because in optics, namely, the intensity of electromagnetic field
plays a main role, which is determined via average vector Poynting
$<\vec {S}>$:

\begin{equation}
<\vec {S}>=\frac{c}{4\pi} \int \biggl(\vec {E}(\vec {r},t)\times
\vec {H}(\vec {r},t)\biggl)dV
\end{equation}
The, we may determine the Hamiltonian of the radiation field
$\hat{H}_R$ in homogeneous medium by following formulae:

\begin{equation}
\hat{H}_R=\frac{1}{8\pi} \int \biggl(\varepsilon^{\frac{1}{2}}(\vec
{r},,\vec{r}^{'};t)E^2+\frac{H^2}{\varepsilon^{\frac{1}{2}}(\vec
{r},\vec{r}^{'};t)}\biggl) dV
\end{equation}

For plane electromagnetic wave propagating in direction of unit
vector $\vec {s}$:

\begin{equation}
\vec {H}(\vec {r},t)=\sqrt{\varepsilon(\vec
{r},\vec{r}^{'};t)}\biggl(\vec {s}\times \vec {E}(\vec {r},t)\biggl)
\end{equation}

and

\begin{equation}
\vec {E}(\vec {r},t)=-\frac{1}{\sqrt{\varepsilon(\vec
{r},\vec{r}^{'};t)}}\biggl(\vec {s}\times \vec {H}(\vec
{r},t)\biggl)
\end{equation}

which substituting into (92), in turn we get the result

\begin{equation}
<\vec {S}>=2c \hat{H}_R \vec {s}
\end{equation}

where wave normal $\vec {s}$ gives a direction propagation of plane
wave, and in turn direction of the average vector Poynting $<\vec
{S}>$.

Thus, we postulate that the Hamiltonian of the radiation field in
homogeneous medium for radiation, propagating toward fixed direction
$\vec {s}$, is presented by (93) but not by

\begin{equation}
\hat{H}_{R,h} =\frac{1}{8\pi V}\int \biggl(\vec {E}\cdot \vec {D}+\mu
H^2\biggl) dV
\end{equation}

which is calculated by case of dielectric respond $\varepsilon(\vec
{r},\vec{r}^{'};t)$ is the constant.

Now, we calculate the radiation Hamiltonian $\hat{H}_R$ in
homogeneous medium (93):

\begin{equation}
\begin{array}{ll}
\hat{H}_R =\frac{1}{8\pi} \int \biggl(\varepsilon^{\frac{1}{2}}(\vec {r},t)E^2+\frac{H^2}{\varepsilon^{\frac{1}{2}}(\vec {r},t)}\biggl) dV =\\[+12pt]
\displaystyle
=\;\frac{1}{8\pi}\int \biggl [\varepsilon^{\frac{1}{2}}(\vec {r},t)\biggl(- \frac{\alpha}{c}\frac{\partial {\vec{H}_0(\vec {r},t)}}{\partial t}+
\beta \vec {E}_0(\vec {r},t)\biggl)^2+\\[+12pt]
\displaystyle +\;\frac{1}{\varepsilon^{\frac{1}{2}}(\vec {r},t)}\biggl(\frac{\alpha}{c}\frac{\partial
{\vec{D}_0}(\vec {r},t)}{\partial t}+ \beta \vec {H}_0(\vec
{r},t)\biggl)^2\biggl] dV
\end{array}
\end{equation}

After simple calculation, in analogy manner as it was made in above,
with using of approximation of a macroscopic number of Light-
Particles in the condensate $N_{0,h}\gg 1$, the radiation
Hamiltonian $\hat{H}_R$ in (98) may rewrite down as

\begin{equation}
\begin{array}{ll}
\hat{H}_R
=\frac{\varepsilon^{\frac{1}{2}}(\vec{k}=0,\omega_{\vec{k}=0})\beta^2
A^2 N_{0,h}}{2\pi}+\sum_{0< k< k_0}\varepsilon(^{\frac{1}{2}}\vec
{k},\omega_{\vec{k}})\biggl [\biggl (\frac{\hbar^2 k^2 }{2m }+
\frac{\beta^2 A^2}{4\pi}\biggl )
\vec{b}^{+}_{\vec{k}}\vec{b}_{\vec{k}}-\\[+12pt]
\displaystyle -\;\frac{1}{2}\biggl (\frac{\hbar^2 k^2 }{2m }-
\frac{\beta^2 A^2}{4\pi}\biggl )\biggl (\vec{b}^{+}_{\vec{k}}
\vec{b}^{+}_{-\vec{k}}+
\vec{b}_{-\vec{k}}\vec{b}_{\vec{k}}\biggl)\biggl ]+\\[+12pt]
\displaystyle +\;\sum_{0< k^{'}< k_0}\varepsilon^{\frac{1}{2}}(\vec
{k}^{'},\omega_{\vec{k}^{'}})\biggl[\biggl(\frac{\hbar^2 {k^{'}}^2
}{2m }+ \frac{\beta^2 A^2}{4\pi}\biggl )
\vec{b}^{+}_{\vec{k}^{'}}\vec{b}_{\vec{k}^{'}}+\\[+12pt]
\displaystyle +\;\frac{1}{2}\biggl (\frac{\hbar^2 {k^{'}}^2 }{2m }-
\frac{\beta^2 A^2}{4\pi}\biggl ) \biggl (\vec{b}^{+}_{\vec{k}^{'}}
\vec{b}^{+}_{-\vec{k}^{'}}+
\vec{b}_{-\vec{k}^{'}}\vec{b}_{\vec{k}}^{'}\biggl)\biggl ]
\end{array}
\end{equation}

Obviously, the operator Hamiltonian $\hat{H}_R$ in (99) takes a
diagonal form by using of new transformation for vector-operator
[1], for evaluation the energy levels of the operator $\hat{H}_R$:
\begin{equation}
\vec{b}_{\vec{k}}=\frac{\vec{d}_{\vec{k}} +
M_{\vec{k}}\vec{d}^{+}_{-\vec{k}}} {\sqrt{1-M^2_{\vec{k}}}}
\end{equation}

where $M_{\vec{k}}$ is the real symmetrical
functions of a wave vector $\vec{k}$.

Then, we find the finally form of $\hat{H}_R$:

\begin{equation}
\hat{H}_R=\hat{H}_{R,0}+2\sum_{ 0< k<
k_0}E_{\vec{k}}\vec{d}^{+}_{\vec{k}}\vec{d}_{\vec{k}}
\end{equation}

where term
\begin{equation}
\hat{H}_{R,0}=mc^2N_{0,h}\varepsilon^{\frac{1}{2}}(\vec{k}=0,\omega_{\vec{k}=0})
\end{equation}

is the energy of Light Particles in the condensate. Otherwise, the
Light Particles in the condensate represent as the constant electric
field, obviously without magnetic one.

Hence, we infer that the Bose-operators $\hat{d}^{+}_{\vec{k}}$ and
$\hat{d}_{\vec{k}}$ are, respectively, the vector creation and
annihilation operators of free polaritons with energy

\begin{equation}
E(\vec {k},\omega_{\vec{k}}) =\hbar k c\varepsilon^{\frac{1}{2}}(\vec
{k},\omega_{\vec{k}})=\frac{\hbar k^2 c^2}{\omega_{\vec{k}}}
\end{equation}
where taking account Eq.(88).

Thus, the energy of polariton $E(\vec {k},\omega_{\vec{k}})$, as
function of $\vec {k}$ and $\omega_{\vec{k}}$ simultaneously, is
presented by (103). However, it needs to present $E(\vec
{k},\omega_{\vec{k}})$ as function of only a wave number $\vec {k}$.
To solve this problem, we investigate the thermodynamical and
optical properties of metal.

\vspace{5mm}

{\bf V. Thermodynamical and optical properties of metal.}

\vspace{5mm}

As we had been showed in recent our paper [18], the formation of a
free neutron singlet pairs in a superfluid liquid helium-dilute
neutron gas mixture which occurs by term, of the interaction between
the excitations of the Bose gas and the density modes of the
neutron, mediate an attractive interaction via the neutron modes,
which in turn leads to a bound state on a spinless neutron pair.
After, we investigated the problem of superconductivity [19]
presented by Fr$\ddot o$lich [20], and demonstrated that the
Fr$\ddot o$lich charged spinless electron pair is created in a
phonon gas-electron gas mixture by the term of the interaction
between the phonon excitations and electron modes by an induced the
effective attractive interaction via electron modes. In turn, there
is a bound state on singlet electron pair with binding energy

$$
E_0
=-\sqrt{\frac{\alpha\hbar^2}{m_e}\biggl(\frac{n}{V}\biggl)^{\frac{5}{3}}}+\frac{\alpha
n }{V} <0
$$
which provides the formation of the superconducting phase in
superconductor by condition for density electrons of metal
$\frac{n}{V}$:
$$
\frac{n}{V}>\biggl(\frac{C^2 m_e }{2M
s^2\hbar^2}\biggl)^{\frac{3}{2}}
$$
At choosing $C\approx 1ev$ [19]; $M\approx 5\cdot 10^{-26} kg$;
$s\approx 3\cdot 10^3 \frac{m}{sec}$, we estimated that any metal
with density of electrons $\frac{n}{V}>10^{27}m^{-3}$ can represent
as a superconductor. This assumption is able to explain the isotope
effect [21]. However, the strong enough lattice-electron interaction
can account the Meissner-Ochsenfeld effect [22] which shows that the
superconductor in the presence of an applied magnetic field, at
temperature below transition temperature superconductor, leads to an
expulsion of the field from the superconductor. In this respect, we
suggest that the Meissner-Ochsenfeld effect may be defined by BEC of
Light Particles in medium which are been by action of the applied
magnetic field. This reasoning implies that the Meissner-Ochsenfeld
effect is not connected with transition temperature superconductor
but it may be defined by transition temperature of the Bose-gas
consisting of the Light Particles. Indeed, the condensed Light
Particles in homogeneous medium is not having a magnetic field,
which follows from (89).

Now, we may remark the most successful attempt made by Schafroth
[23], who considered a model superconductor as the charged ideal
Bose gas, consisting of charged electron pairs with charge $e_0=2e$
and mass $m_0=2 m_e$. He defined a superconducting phase in the
superconductor by using of a density of charged bosons in the
condensate. We call these singlet electron pairs by the names of the
Fr$\ddot o$lich and Schafroth because first Schafroth stated the
existence these singlet electron pairs into superconductor with
charge $e_0=2e$ and mass $m_0=2 m_e$ but Fr$\ddot o$lich could
discover their earlier then it was made by the Cooper. Thus, any
metal may consider by the model of free Fr$\ddot o$lich-Schafroth's
charged singlet bosons with charge $e_0=2e$ and mass $m_0=2 m_e$, if
the density of electrons of metal satisfies to the condition
$\frac{n}{V}>10^{27}m^{-3}$. Otherwise, if metal consists an ion
lattice and electron gas, then, the free Fr$\ddot o$lich-Schafroth's
charged singlet bosons are formed.

Hence, we note that the Fr$\ddot o$lich- Schafroth charged bosons
are differ from Cupper's pairs [24] which are formed by electrons
filling near to the Fermi level. In this respect, we note that
theory of superconductivity, presented by Bardeen, Cooper and
Schrieffer, and by Bogoliubov (BCSB) [25] based on the existence of
the Cupper's pairs. They asserted that the Fr$\ddot o$lich effective
attractive potential between electrons leads to shaping of two
electrons with opposite spins to so-called Cooper pairs around Fermi
level. However, hence we suggests that their theory contradicts to
the principle of identity of particles because the same electrons
can not be separated by free electrons into Fermi levels and by
interacting electrons around Fermi one, as they state the Paul
exclusion principle. However, this principle states that two
identical fermions (particles with half-integer spin) may occupy the
same quantum state simultaneously. In fact, the Paul principle is
not connected with interaction between fermions as it was accepted
in the theory of BCSB.

It is well known that the one-particle excitation spectrum in the
charged Bose gas was first investigated by Foldy [26] at $T=0$ and
by Bishop [27] at or near transition temperature $T_c$. The
random-phase approximation (RPA) dielectric response for ideal
charged Bose gas, at finite temperatures, was proposed by  Hore and
Frenkel [28], where was taken the Coulomb interaction between
charged particles. On other hand, as we have been mentioned in
above, any metal may consider by the model of an ideal Bose gas
consisting of free Fr$\ddot o$lich-Schafroth's charged singlet
bosons with charge $e_0=2e$ and mass $m_0=2 m_e$, if the density
electrons of metal satisfies to the condition
$\frac{n}{V}>10^{27}m^{-3}$. In accordance this fact, we use of the
Schafroth's model of an ideal charged Bose-gas [19] where the
density of bosons in the condensate $\rho_s$ depends on temperature
$T$:

\begin{equation}
\rho_s=\rho\biggl[1-\biggl(\frac{T}{T_c}\biggl)^{\frac{3}{2}}\biggl]
\end{equation}

where $\rho=\frac{1}{2}\cdot\frac{n}{V}$ is the density of bosons
presented via density of electrons of metal; $\zeta(x)$ is the
Riemann zeta function; $k$ is the Boltzman constant; $T_c$ is the
transition temperature.

At $\rho_s=0$, the transition temperature $T_c$ is
$$
T_c=\frac{2\pi\hbar^2}{m_0 k}\biggl(\frac{\rho}{\zeta
(\frac{3}{2})}\biggl)^{\frac{2}{3}}
$$
For any metal, the density electrons is $\frac{n}{V}\approx
10^{28}m^{-3}$, therefore, at $\rho=0.5\approx 10^{28}m^{-3}$, we
have $T_c\approx 10^3 K$. This result implies that at room
temperature $T=300K$, we may consider any metal as superconductor
because $T\leq T_c$.

Otherwise, the clear form of the complex dielectric respond
$\hat{\varepsilon}(\vec {k},\omega)$, as function from temperature
$T$ for ideal charged Bose gas [28] which at the temperatures $T\leq
T_c$ is presented by the form:

\begin{equation}
\begin{array}{ll}
\hat{\varepsilon}(\vec {k},\omega_{\vec {k}})=1-\frac{\omega^2_p}{\omega^2_{\vec {k}}-\frac{\hbar^2
k^4}{4m^2_0}}\biggl[1-\biggl(\frac{T}{T_c}\biggl)^{\frac{3}{2}}\biggl]
+\\[+12pt]
\displaystyle +\;\frac{\omega^2_p}{\rho}\cdot \frac{1}{4\pi^2}\cdot
\frac{m^2_0}{\hbar^2k^3}\cdot\frac{2m_0 kT}{\hbar^2}\frac{\pi}{2i}\times\\[+12pt]
\displaystyle \times\;\sum^{\infty}_{j=1}\frac{1}{j}\biggl[\biggl(
1+\phi(DC_j)\biggl)e^{D^2C^2_j}-\biggl(
1+\phi(BC_j)\biggl)e^{B^2C^2_j} \biggl]
\end{array}
\end{equation}

where $D=i\biggl(\frac{m_0\omega_{\vec {k}}}{\hbar k}+\frac{1}{2}k\biggl)$, $B=i\biggl(\frac{m_0\omega_{\vec {k}}}{\hbar k}-\frac{1}{2}k\biggl)$, $C_j=A^{\frac{1}{2}}j^{\frac{1}{2}}$; $A=\frac{\hbar^2}{2m_0 kT}$; $\Phi(x)$ is the error function:
$$
\Phi(x)=\frac{2}{\sqrt{\pi}}\int^{x}_{0}e^{-t^2}dt
$$

and $\omega^2_p=\frac{4\pi e^2_0\rho}{m_0}$ is the plasma frequency for the charged gas.

After some algebra, they found the following expansions for
$\hat{\varepsilon}(\vec {k},\omega_{\vec {k}})$ in (93) at $T\leq
T_c$, which may present by the form [14]:

\begin{equation}
\hat{\varepsilon}(\vec {k},\omega_{\vec {k}})=\varepsilon(\vec {k},\omega_{\vec {k}})+i\frac{4\pi\sigma(\vec {k},\omega)}{\omega}
\end{equation}

where

\begin{equation}
\begin{array}{ll}
\varepsilon(\vec {k},\omega)=1-\frac{\omega^2_p}{\omega^2_{\vec {k}}-\frac{\hbar^2
k^4}{4m^2_0}}\biggl[1-\biggl(\frac{T}{T_c}\biggl)^{\frac{3}{2}}\biggl]
+\\[+12pt]
\displaystyle +\;\frac{k^2 kT}{m_0}\cdot\frac{\omega^2_p}
{\biggl(\omega^2_{\vec {k}}-\frac{\hbar^2 k^4}{4m^2_0}\biggl)^3}
\frac{\zeta(\frac{5}{2})}{\zeta
(\frac{3}{2})}\cdot\biggl(3\omega^2_{\vec {k}}+\frac{\hbar^2
k^4}{4m^2_0}\biggl)\biggl(\frac{T}{T_c}\biggl)^{\frac{3}{2}}+\cdots+
\end{array}
\end{equation}

\begin{equation}
\begin{array}{ll}
\frac{4\pi\sigma(\vec {k},\omega_{\vec {k}})}{\omega_{\vec {k}}}=\frac{\omega^2_p}{\rho}\cdot
\frac{m^3_0 kT}{3\pi\hbar^4k^3}\biggl(sinh\biggl(\frac{\hbar\omega}{2m_0 kT}\biggl)
\exp^{-\biggl[\biggl(\frac{m_0\omega}{\hbar k}\biggl)^2 +\frac{k^2}{4}\biggl]}\frac{\hbar^2}{2m_0 kT}+\\[+12pt]
\displaystyle +\;\cdots+\biggl)
\end{array}
\end{equation}

hence $\varepsilon(\vec {k},\omega_{\vec {k}})$ is the dielectric constant;
$\sigma(\vec {k},\omega_{\vec {k}})$ is the conductivity of metal. Now, we
introduce the complex refractive index
$\hat{n}=\sqrt{\hat{\varepsilon}(\vec {k},\omega_{\vec {k}})}$ where $n$ is the
real refractive index and $k$ is the attenuation index of metal:

\begin{equation}
\hat{n}=n(1+ik)
\end{equation}

where
\begin{equation}
n^2=\frac{1}{2}\biggl(\sqrt{\varepsilon^2(\vec {k},\omega_{\vec {k}})+\frac{16\pi^2\sigma^2(\vec {k},\omega)}{\omega^2}}+\varepsilon(\vec {k},\omega_{\vec {k}})\biggl)
\end{equation}

and

\begin{equation}
n^2k^2=\frac{1}{2}\biggl(\sqrt{\varepsilon^2(\vec {k},\omega_{\vec {k}})+
\frac{16\pi^2\sigma^2(\vec {k},\omega_{\vec {k}})}{\omega^2_{\vec {k}}}}-\varepsilon(\vec {k},\omega_{\vec {k}})\biggl)
\end{equation}

\vspace{5mm}

{\bf VI. Polaritons in metal and BEC.}

\vspace{5mm}

Now, we show that the equations (88) and (105) can present two
original types of the BEC. Hence, we consider two interesting cases.

1. At  $T=0$, the dielectric response in Eq.(105) takes the form:

\begin{equation}
\varepsilon(\vec {k},\omega_{\vec {k}}, T=0)=1-\frac{\omega^2_p}{\omega^2_{\vec {k}}-\frac{\hbar^2
k^4}{4m^2_0}}
\end{equation}

In the case of the plasmon excitations $\varepsilon(\vec
{k},\omega_{\vec {k},p})=0$, which in turn allows us to find the
frequency of plasmon field:
\begin{equation}
\omega_{\vec {k},p}=\sqrt{\omega^2_p +\frac{\hbar^2
k^4}{4m^2_0}}
\end{equation}
Obviously, at $\varepsilon(\vec {k},\omega_{\vec {k},p})=0$, the
Eq.(88) takes the form:
\begin{equation}
\varepsilon(\vec {k},\omega_{\vec {k},p}, T=0)=\frac{k^2
c^2}{\omega_{\vec {k},p}}
\end{equation}
which is fulfilled at $k\rightarrow 0$, namely, at existence of BEC
because all Light Particles fill in the condensate level $\vec
{k}=0$, and reproduce a constant electric field. At approximation
$k\rightarrow 0$, the dielectric response around the plasmon
frequency $\omega_{\vec {k}}\sim\omega_p$ takes the form
\begin{equation}
\varepsilon(\vec {k},\omega_{\vec {k}})=\frac{k^2 c^2}{\omega_p}
\end{equation}

which is inserting into (103), and, then we obtain the energy
polariton-plasmon field $E(\vec{k},\omega_{\vec{k}}; T=0)$ as the
function of only the wave vector $\vec {k}$:

\begin{equation}
E(\vec{k},\omega_{\vec{k}}; T=0)=\frac{\hbar k^2
c^2}{\omega_p}=\frac{\hbar^2 k^2}{2m_p}
\end{equation}

This result means that the polariton of plasmon field represents as
the Bose-quasiparticle with effective mass

\begin{equation}
m_p=\frac{\hbar\omega_p}{2c^2}\approx 5\cdot10^{-6}m_e
\end{equation}

because for many metals $\omega_p\approx 10^{16}Hz$. Thus, we may
state that near the frequency of plasmon field, the electromagnetic
wave in the metal induces the plasmon-polariton Bose-quasiparticles.

On other hand, we may show that such plasmon-polaritons are able to
be exited into electron gas at absolute zero because at high
frequency limit [29], the dielectric constant of electron gas is

\begin{equation}
\varepsilon=1-\frac{\omega^2_p}{\omega^2_{\vec{k}}}
\end{equation}

In the case of plasmon excitations $\varepsilon=0$, and then, the
frequency of plasmon-polariton $\omega_{\vec{k}}=\omega_p$, which in
turn, determines the energy of plasmon-polariton presented by
Eq.(116).

Thus, the Eq.(116) is an universal because it is a presence at
describing of the Bose gas and the electron gas where the plasmon
frequency $\omega_p$ is not changed in the case of the Bose gas as
well as the electron gas:
$$
\omega_p=\sqrt{\frac{4\pi
e^2_0\rho}{m_0}}=\sqrt{\frac{4\pi e^2 n}{V m_e}}
$$
where $m_0=2m_e$, $e_0=2e$ and $\rho=\frac{1}{2}\cdot\frac{n}{V}$.

Indeed, the quantity of the plasma frequency depends on the density
of electron into metal which may be determined  by gas parameter
$r_s=\frac{m_e e^2}{\hbar^2\biggl(\frac{4\pi
n}{3V}\biggl)^{\frac{1}{3}}}$. This parameter takes a quantity into
interval $1.8\leq r_s\leq 5.5$ for metals [30]. At a quantity of the
plasma frequency $\omega_{p}\approx 10^{16}Hz$, we obtain $m_p$
presented in (117).

2. Now, we consider the case of the resonance effect, when a
frequency of incoming photon $\omega_{\vec {k}}$ coincides with the
frequency of one Fr$\ddot o$lich- Schafroth's charged singlet bosons
$\frac{\hbar k^2}{2m_0}$. In this case, the complex dielectric
respond $\hat{\varepsilon}(\vec {k},\omega_{\vec{k},r})$, presented
by (105), is infinity or $\hat{\varepsilon}(\vec
{k},\omega_{\vec{k},r})=\infty$, at finite temperatures. Thus, the
resonance frequency is

\begin{equation}
\omega_{\vec {k},r}=\frac{\hbar k^2}{2m_0}
\end{equation}

To find frequency of new resonance-polaritons around resonance
frequency $\omega_{\vec {k},r}$, we substitute $\omega_{\vec {k},r}$
into right side of (88) which may determine a dielectric respond for
Light Particles $\varepsilon(\vec{k},\omega_{\vec{k}}; T)$ around
$\omega_{\vec{k},r}$:

\begin{equation}
\sqrt{\varepsilon(\vec{k},\omega_{\vec{k}}; T)}=\frac{2m_0c}{\hbar
k}
\end{equation}

As we see in (121), at $k=0$ follows that $\hat{\varepsilon}(\vec
{k},\omega)=\infty$, which implies the appearance of an original
type BEC, when all Light Particles fill the condensate level $\vec
{k}=0$. On other hand, we may find the energy of
resonance-polaritons $E(\vec{k},\omega_{\vec{k}}; T)$ around
$\omega_{\vec {k},r}$ by substituting $\omega_{\vec {k},r}$ from
(120) into (103), and then,

\begin{equation}
E(\vec{k},\omega_{\vec{k}}; T) =2m_0 c^2
\end{equation}
which does not depend on wave vector $\vec {k}$ and temperature $T$.
Thus,

\begin{equation}
\hat{H}_R=\hat{H}_{R,0}+2\sum_{ 0< k<
k_0}E_{\vec{k}}\vec{d}^{+}_{\vec{k}}\vec{d}_{\vec{k}}
\end{equation}

where the energy of Light Particles in the condensate is
\begin{equation}
\hat{H}_{R,0}=\lim_{k\rightarrow 0}\frac{2m_0 m c^3N_{0,h}}{\hbar
k}\rightarrow \infty
\end{equation}

This result confirms that namely BEC of Light Particles in the
condensate in metal may enhance the optical property of Light and
reproduce the constant electric field.

Thus, the resonance effect around the resonance frequency
$\omega_{\vec {k},r}$ induces the neutral resonance-polariton as
Bose-quasiparticles with a constant energy
$E(\vec{k},\omega_{\vec{k}}; T)=2m_0 c^2$ into metal.

\vspace{5mm}

{\bf VII. Reflection and transmission of plane wave.}

\vspace{5mm}

Now we treat the quantization Fresnel's equations. If an incident
plane wave propagates, by direction of unit vector $\vec {s}^{i}$,
across the interface between air and metal, the intensity of the
wave will be divided between a reflected, by direction of wave
normal $\vec {s}^{r}$, and a refracted, by direction of unit vector
$\vec {s}^{t}$, waves. The vectors electric and magnetic fields of
incident $\vec{E}_{0,i} (\vec{r}, t)$ and $\vec{H}_{0,i} (\vec{r},
t)$, reflected $\vec{E}_{0,r}(\vec{r}, t)$ and $\vec{H}_{0,r}
(\vec{r}, t)$, and refracted $\vec{E}_{0,t} (\vec{r}, t)$ and
$\vec{H}_{0,t} (\vec{r}, t)$ Light-Particles  by the matrices:

\begin{equation}
\begin{array}{ll}
\vec{E}_{0,i}=\left(
\begin{array}{ccc}\vec{E}^{i}_x\\\vec{E}^{i}_y\\\vec{E}^{i}_z
\end{array}
\right)=\\[+12pt]\displaystyle
=\;\frac{1}{\sqrt{V}}\sum_{\vec{k}}\left(
\begin{array}{ccc}-\vec{a}_{\vec{k},{\mid\mid}}cos\theta_i\\\vec{a}_{\vec{k},{\bot}}\\\vec{a}_{\vec{k},{\mid\mid}}sin\theta_i
\end{array}
\right)e^{i\tau_i}+ \frac{1}{\sqrt{V}}\sum_{\vec{k}}\left(
\begin{array}{ccc}-\vec{a}^{+}_{\vec{k},{\mid\mid}}cos\theta_i\\\vec{a}^{+}_{\vec{k},{\bot}}\\\vec{a}^{+}_{\vec{k},{\mid\mid}}sin\theta_i
\end{array}
\right)e^{-i\tau_i}
\end{array}
\end{equation}

\begin{equation}
\begin{array}{ll}
\vec{E}_{0,r}=\left(
\begin{array}{ccc}\vec{E}^{r}_x\\\vec{E}^{r}_y\\\vec{E}^{r}_z
\end{array}
\right)=\\[+12pt]\displaystyle
=\;\frac{1}{\sqrt{V}}\sum_{\vec{k}}\left(
\begin{array}{ccc}-\vec{c}_{\vec{k},{\mid\mid}}cos\theta_r\\\vec{c}_{\vec{k},{\bot}}\\\vec{c}_{\vec{k},{\mid\mid}}sin\theta_r
\end{array}
\right)e^{i\tau_r}+\frac{1}{\sqrt{V}}\sum_{\vec{k}}\left(
\begin{array}{ccc}-\vec{c}^{+}_{\vec{k},{\mid\mid}}cos\theta_r\\\vec{c}^{+}_{\vec{k},{\bot}}\\\vec{c}^{+}_{\vec{k},{\mid\mid}}sin\theta_r
\end{array}
\right)e^{-i\tau_r}
\end{array}
\end{equation}

\begin{equation}
\begin{array}{ll}
\vec{E}_{0,t}=\left(
\begin{array}{ccc}\vec{E}^{t}_x\\\vec{E}^{t}_y\\\vec{E}^{t}_z
\end{array}
\right)=\\[+12pt]\displaystyle
=\;frac{1}{\sqrt{V}}\sum_{\vec{k}}\left(
\begin{array}{ccc}-\vec{b}_{\vec{k},{\mid\mid}}cos\theta_t\\\vec{b}_{\vec{k},{\bot}}\\\vec{b}_{\vec{k},{\mid\mid}}sin\theta_t
\end{array}
\right)e^{i\tau_t}+\frac{1}{\sqrt{V}}\sum_{\vec{k}}\left(
\begin{array}{ccc}-\vec{b}^{+}_{\vec{k},{\mid\mid}}cos\theta_t\\\vec{b}^{+}_{\vec{k},{\bot}}\\\vec{b}^{+}_{\vec{k},{\mid\mid}}sin\theta_t
\end{array}
\right)e^{-i\tau_t}
\end{array}
\end{equation}

and using (94), we have
\begin{equation}
\begin{array}{ll}
\vec{H}_{0,i}=\left(
\begin{array}{ccc}\vec{H}^{i}_x\\\vec{H}^{i}_y\\\vec{H}^{i}_z
\end{array}
\right)=\\[+12pt]\displaystyle
=\;\frac{1}{\sqrt{V}}\sum_{\vec{k}}\left(
\begin{array}{ccc}-\vec{a}_{\vec{k},\bot}cos\theta_i\\\vec{a}_{\vec{k},{\mid\mid}}\\\vec{a}_{\vec{k},\bot}sin\theta_i
\end{array}
\right)e^{i\tau_i}+ \frac{1}{\sqrt{V}}\sum_{\vec{k}}\left(
\begin{array}{ccc}-\vec{a}^{+}_{\vec{k},\bot}cos\theta_i\\\vec{a}^{+}_{\vec{k},{\mid\mid}}\\\vec{a}^{+}_{\vec{k},\bot}sin\theta_i
\end{array}
\right)e^{-i\tau_i}
\end{array}
\end{equation}

\begin{equation}
\begin{array}{ll}
\vec{H}_{0,r}=\left(
\begin{array}{ccc}\vec{H}^{r}_x\\\vec{H}^{r}_y\\\vec{H}^{r}_z
\end{array}
\right)=\\[+12pt]\displaystyle
=\;frac{1}{\sqrt{V}}\sum_{\vec{k}}\left(
\begin{array}{ccc}-\vec{c}_{\vec{k},\bot}cos\theta_i\\\vec{c}_{\vec{k},{\mid\mid}}\\\vec{c}_{\vec{k},\bot}sin\theta_i
\end{array}
\right)e^{i\tau_r}+ \frac{1}{\sqrt{V}}\sum_{\vec{k}}\left(
\begin{array}{ccc}-\vec{c}^{+}_{\vec{k},\bot}cos\theta_i\\\vec{c}^{+}_{\vec{k},{\mid\mid}}\\\vec{c}^{+}_{\vec{k},\bot}sin\theta_i
\end{array}
\right)e^{-i\tau_r}
\end{array}
\end{equation}
and

\begin{equation}
\begin{array}{ll}
\vec{H}_{0,t}=\left(
\begin{array}{ccc}\vec{H}^{t}_x\\\vec{H}^{t}_y\\\vec{H}^{t}_z
\end{array}
\right)=\frac{1}{\sqrt{V}}\sum_{\vec{k}}\left(
\begin{array}{ccc}-\sqrt{\varepsilon(\vec
{k},\omega_{\vec{k}})}\vec{b}_{\vec{k},\bot}cos\theta_i\\\sqrt{\varepsilon(\vec
{k},\omega_{\vec{k}})}\vec{b}_{\vec{k},{\mid\mid}}\\\vec{b}_{\vec{k},\bot}sin\theta_i
\end{array}
\right)e^{i\tau_t}+ \\[+12pt]\displaystyle
+\;\frac{1}{\sqrt{V}}\sum_{\vec{k}}\left(
\begin{array}{ccc}-\sqrt{\varepsilon(\vec
{k},\omega_{\vec{k}})}\vec{b}^{+}_{\vec{k},\bot}cos\theta_i\\\sqrt{\varepsilon(\vec
{k},\omega_{\vec{k}})}\vec{b}^{+}_{\vec{k},{\mid\mid}}\\\vec{b}^{+}_{\vec{k},\bot}sin\theta_i
\end{array}
\right)e^{-i\tau_t}
\end{array}
\end{equation}

where
$$
\tau_i= \vec{k}\vec{r} + \omega_{\vec{k}}
t=\omega_{\vec{k}}\biggl(\frac{\vec{s}_i\cdot\vec{r}}{c}+t\biggl)
$$

$$
\tau_r= \vec{k}\vec{r} + \omega_{\vec{k}}
t=\omega_{\vec{k}}\biggl(\frac{\vec{s}_r\cdot\vec{r}}{c}+t\biggl)
$$

and
$$
\tau_t= \vec{k}\vec{r} + \omega_{\vec{k}}
t=\omega_{\vec{k}}\biggl(\frac{\sqrt{\varepsilon(\vec
{k},\omega_{\vec{k}})}\vec{s}_t\cdot\vec{r}}{c}+t\biggl)
$$

are, respectively, the arguments of the incident, reflected and
refracted Light Particles in the space wave vector $\vec{k}$ and
frequency $\omega_{\vec{k}}$, where
$\omega_{\vec{k}}=\frac{kc}{\sqrt{\varepsilon(\vec
{k},\omega_{\vec{k}})}}$ (88).

Due to Huygens's principle, every point in direction of the Light
represents as the source of second waves, therefore, in point of
coordinate $\vec{r}=\biggl(x, y, 0\biggl)$ on the threshold plane
$z=0$ at interface between air and metal in the coordinate system
$XYZ$, the arguments of wave functions are the same
$\tau_i=\tau_r=\tau_t$ [22]. Consequently, the laws of refraction
and reflection on the boundary air-metal, takes the form:
\begin{equation}
sin\theta_t= \frac{sin\theta_i}{\hat{n}(\vec {k},\omega_{\vec{k}})}
\end{equation}

and

\begin{equation}
\theta_t= \pi-\theta_r
\end{equation}

where $\theta_i$, $\theta_r$ and $\theta_t$ are, respectively, the
angles of incident, reflection and refraction;
$\hat{n}=\sqrt{\varepsilon(\vec {k},\omega_{\vec{k}})}$.

Considering the vector Bose-operators $\vec{a}_{\vec{k}}$,
$\vec{c}_{\vec{k}}$ and $\vec{b}_{\vec{k}}$ are, respectively,
expanded to components of the normal $\vec{a}_{\vec{k},{\bot}}$,
$\vec{c}_{\vec{k},{\bot}}$, $\vec{b}_{\vec{k},{\bot}}$ and parallel
to the plane of incidence $\vec{a}_{\vec{k},{\mid\mid}}$,
$\vec{c}_{\vec{k},{\mid\mid}}$, $\vec{b}_{\vec{k},{\mid\mid}}$,
then, amplitudes of the wave are determined by the boundary
conditions requiring that the tangential components of vectors
electric and magnetic fields must be the same on both sides of
interface air-metal:

\begin{equation}
\left.
\begin{array}{c}
\vec{E}^{i}_x+\vec{E}^{r}_x=\vec{E}^{t}_x \\
\vec{E}^{i}_y+\vec{E}^{r}_y=\vec{E}^{t}_y\\
\vec{H}^{i}_x+\vec{H}^{r}_x=\vec{H}^{t}_x \\
\vec{H}^{i}_y+\vec{H}^{r}_y=\vec{H}^{t}_y\\
\end{array} \right\}
\end{equation}

These boundary conditions lead to the relationships between
operators $\vec{c}_{\vec{k},{\bot}}$, $\vec{b}_{\vec{k},{\bot}}$ and
$\vec{c}_{\vec{k},{\mid\mid}}$, $\vec{b}_{\vec{k},{\mid\mid}}$ via
$\vec{a}_{\vec{k},{\bot}}$ and $\vec{a}_{\vec{k},{\mid\mid}}$.
Indeed, letting that the vector incidence electric field $\vec
E_{0,i}$ has an angle $\alpha_i$ with the plane of incidence, then
we have
\begin{equation}
\left.
\begin{array}{c}
\vec{a}_{\vec{k},{\bot}}=\vec{a}_{\vec{k}}sin\alpha_i \\
\vec{a}_{\vec{k},{\mid\mid}}=\vec{a}_{\vec{k}}cos\alpha_i\\
\end{array}
\right\}
\end{equation}

which allows us to find

\begin{equation}
\left.
\begin{array}{c}
\vec{c}_{\vec{k},{\bot}}=r_{\bot}\vec{a}_{\vec{k},{\bot}}=r_{\bot}\vec{a}_{\vec{k}}sin\alpha_i \\
\vec{c}_{\vec{k},{\mid\mid}}=r_{\mid\mid}\vec{a}_{\vec{k},{\mid\mid}}=r_{\mid\mid}\vec{a}_{\vec{k}}cos\alpha_i\\
\vec{c}^{+}_{\vec{k},{\bot}}=r^{\ast}_{\bot}\vec{a}^{+}_{\vec{k},{\bot}}=r^{\ast}_{\bot}\vec{a}^{+}_{\vec{k}}sin\alpha_i \\
\vec{c}^{+}_{\vec{k},{\mid\mid}}=r^{\ast}_{\mid\mid}\vec{a}^{+}_{\vec{k},{\mid\mid}}=r^{\ast}_{\mid\mid}\vec{a}^{+}_{\vec{k}}cos\alpha_i
\end{array}
\right\}
\end{equation}

and

\begin{equation}
\left.
\begin{array}{c}
\vec{b}_{\vec{k},{\bot}}=t_{\bot}\vec{a}_{\vec{k},{\bot}}=t_{\bot}\vec{a}_{\vec{k}}sin\alpha_i \\
\vec{b}_{\vec{k},{\mid\mid}}=t_{\mid\mid}\vec{a}_{\vec{k},{\mid\mid}}=t_{\mid\mid}\vec{a}_{\vec{k}}cos\alpha_i\\
\vec{b}^{+}_{\vec{k},{\bot}}=t^{\ast}_{\bot}\vec{a}^{+}_{\vec{k},{\bot}}=t^{\ast}_{\bot}\vec{a}^{+}_{\vec{k}}sin\alpha_i \\
\vec{b}_{\vec{k},{\mid\mid}}=t^{\ast}_{\mid\mid}\vec{a}^{+}_{\vec{k},{\mid\mid}}=t^{\ast}_{\mid\mid}\vec{a}^{+}_{\vec{k}}cos\alpha_i
\end{array}
\right\}
\end{equation}

where quantities $r_{\bot}$, $r_{\mid\mid}$ and $t_{\bot}$,
$t_{\mid\mid}$ are the amplitudes of reflection and refraction
coefficients of normal and parallel to the plane of incidence,
respectively, which present the Fresnel's equations:

\begin{equation}
r_{\bot}=\frac{cos\theta_i -\hat{n}(\vec {k},\omega_{\vec{k}})
cos\theta_t}{cos\theta_i +\hat{n}(\vec {k},\omega_{\vec{k}})
cos\theta_t}
\end{equation}

\begin{equation}
r_{\mid\mid}=\frac{\hat{n}(\vec {k},\omega_{\vec{k}})cos\theta_i -
cos\theta_t}{\hat{n}(\vec {k},\omega_{\vec{k}})cos\theta_i +
cos\theta_t}
\end{equation}
and
\begin{equation}
t_{\bot}=\frac{2 cos\theta_i }{cos\theta_i +\hat{n}(\vec
{k},\omega_{\vec{k}}) cos\theta_t}
\end{equation}

\begin{equation}
t_{\mid\mid}=\frac{2 cos\theta_i }{\hat{n}(\vec
{k},\omega_{\vec{k}})cos\theta_i + cos\theta_t}
\end{equation}

Now, we try to treat the Hamiltonian operators of incident
$\hat{H}_{R,i}$, reflected $\hat{H}_{R,r}$ and refracted
$\hat{H}_{R,t}$ waves. In this respect, we separate the Hamiltonian
operators. So, we may rewrite the expanded forms of the Hamiltonian
operators of incident $\hat{H}_{R,i}$ radiation by following form:

\begin{equation}
\hat{H}_{R,i}=\hat{H}_{i,{\mid\mid}}+\hat{H}_{i,{\bot}}
\end{equation}

where

\begin{equation}
\begin{array}{ll}
\hat{H}_{i,{\mid\mid}}=mc^2\vec{a}^{+}_{0, {\mid\mid}}\vec{a}_{0,
{\mid\mid}}+\sum_{0<k<k_0}\biggl(\frac{\hbar^2 k^2 }{2m }+
\frac{mc^2}{2}\biggl)
\vec{a}^{+}_{\vec{k},{\mid\mid}}\vec{a}_{\vec{k},{\mid\mid}}-\\[+12pt]
\displaystyle -\;\frac{1}{2}\sum_{0<k<k_0}\biggl(\frac{\hbar^2 k^2
}{2m }- \frac{mc^2}{2}\biggl) \biggl
(\vec{a}^{+}_{\vec{k},{\mid\mid}} \vec{a}^{+}_{-\vec{k},{\mid\mid}}+
\vec{a}_{-\vec{k},{\mid\mid}}\vec{a}_{\vec{k},{\mid\mid}}\biggl) +\\[+12pt]
\displaystyle +\; \sum_{0<k^{'}<k_0}\biggl(\frac{\hbar^2 {k^{'}}^2
}{2m }+ \frac{mc^2}{2}\biggl)
\vec{a}^{+}_{\vec{k}^{'},{\mid\mid}}\vec{a}_{\vec{k}^{'},{\mid\mid}}+\\[+12pt]
\displaystyle
+\;\frac{1}{2}\sum_{0<k^{'}<k^{'}_0}\biggl(\frac{\hbar^2 k^2 }{2m }-
\frac{mc^2}{2}\biggl)\biggl (\vec{a}^{+}_{\vec{k}^{'},{\mid\mid}}
\vec{a}^{+}_{-\vec{k}^{'},{\mid\mid}}+
\vec{a}_{-\vec{k}^{'},{\mid\mid}}\vec{a}_{\vec{k}^{'},{\mid\mid}}\biggl)
\end{array}
\end{equation}

and
\begin{equation}
\begin{array}{ll}
\hat{H}_{i,{\bot}}=mc^2\vec{a}^{+}_{0, {\bot}}\vec{a}_{0,
{\bot}}+\sum_{0<k<k_0}\biggl(\frac{\hbar^2 k^2 }{2m }+
\frac{mc^2}{2}\biggl)
\vec{a}^{+}_{\vec{k},{\bot}}\vec{a}_{\vec{k},{\bot}}-\\[+12pt]
\displaystyle -\;\frac{1}{2}\sum_{0<k<k_0}\biggl(\frac{\hbar^2 k^2
}{2m }- \frac{mc^2}{2}\biggl) \biggl (\vec{a}^{+}_{\vec{k},{\bot}}
\vec{a}^{+}_{-\vec{k},{\bot}}+
\vec{a}_{-\vec{k},{\bot}}\vec{a}_{\vec{k},{\bot}}\biggl) +\\[+12pt]
\displaystyle +\; \sum_{0<k^{'}<k_0}\biggl(\frac{\hbar^2 {k^{'}}^2
}{2m }+ \frac{mc^2}{2}\biggl)
\vec{a}^{+}_{\vec{k}^{'},{\bot}}\vec{a}_{\vec{k}^{'},{\bot}}+\\[+12pt]
\displaystyle
+\;\frac{1}{2}\sum_{0<k^{'}<k^{'}_0}\biggl(\frac{\hbar^2 k^2 }{2m }-
\frac{mc^2}{2}\biggl) \biggl (\vec{a}^{+}_{\vec{k}^{'},{\bot}}
\vec{a}^{+}_{-\vec{k}^{'},{\bot}}+
\vec{a}_{-\vec{k}^{'},{\bot}}\vec{a}_{\vec{k}^{'},{\bot}}\biggl)
\end{array}
\end{equation}

where taking into account (133) with transformation (32), we get

\begin{equation}
\hat{H}_{R,i}=\hat{H}_{0,i}+2\sum_{0< k<
k_0}\chi_{\vec{k}}\vec{i}^{+}_{\vec{k}}\vec{i}_{\vec{k}}
\end{equation}
where

\begin{equation}
\hat{H}_{0,i}=mc^2N_0
\end{equation}

is the energy of Light-Particles of incidence Light in the
condensate;
\begin{equation}
\chi_{\vec{k}}=\hbar k c
\end{equation}
is energy of incoming photons; $\vec{i}^{+}_{\vec{k}}$ and
$\vec{i}_{\vec{k}}$  are, respectively, the Bose vector-operators of
creation and annihilation of photons.

In analogy manner, we may rewrite the Hamiltonian operator of
reflected $\hat{H}_{R,r}$ radiation:
\begin{equation}
\hat{H}_{R,r}=\hat{H}_{r,{\mid\mid}}+\hat{H}_{r,{\bot}}
\end{equation}

where

\begin{equation}
\begin{array}{ll}
\hat{H}_{r,{\mid\mid}}=mc^2\vec{c}^{+}_{0, {\mid\mid}}\vec{c}_{0,
{\mid\mid}}+\sum_{0<k<k_0}\biggl(\frac{\hbar^2 k^2 }{2m }+
\frac{mc^2}{2}\biggl)
\vec{c}^{+}_{\vec{k},{\mid\mid}}\vec{c}_{\vec{k},{\mid\mid}}-\\[+12pt]
\displaystyle -\;\frac{1}{2}\sum_{0<k<k_0}\biggl(\frac{\hbar^2 k^2
}{2m }- \frac{mc^2}{2}\biggl) \biggl
(\vec{c}^{+}_{\vec{k},{\mid\mid}} \vec{c}^{+}_{-\vec{k},{\mid\mid}}+
\vec{c}_{-\vec{k},{\mid\mid}}\vec{c}_{\vec{k},{\mid\mid}}\biggl) +\\[+12pt]
\displaystyle +\; \sum_{0<k^{'}<k_0}\biggl(\frac{\hbar^2 {k^{'}}^2
}{2m }+ \frac{mc^2}{2}\biggl)
\vec{c}^{+}_{\vec{k}^{'},{\mid\mid}}\vec{c}_{\vec{k}^{'},{\mid\mid}}+\\[+12pt]
\displaystyle
+\;\frac{1}{2}\sum_{0<k^{'}<k^{'}_0}\biggl(\frac{\hbar^2 k^2 }{2m }-
\frac{mc^2}{2}\biggl)\biggl (\vec{c}^{+}_{\vec{k}^{'},{\mid\mid}}
\vec{c}^{+}_{-\vec{k}^{'},{\mid\mid}}+
\vec{c}_{-\vec{k}^{'},{\mid\mid}}\vec{c}_{\vec{k}^{'},{\mid\mid}}\biggl)
\end{array}
\end{equation}

and
\begin{equation}
\begin{array}{ll}
\hat{H}_{r,{\bot}}=mc^2\vec{c}^{+}_{0, {\bot}}\vec{c}_{0,
{\bot}}+\sum_{0<k<k_0}\biggl(\frac{\hbar^2 k^2 }{2m }+
\frac{mc^2}{2}\biggl)
\vec{c}^{+}_{\vec{k},{\bot}}\vec{c}_{\vec{k},{\bot}}-\\[+12pt]
\displaystyle -\;\frac{1}{2}\sum_{0<k<k_0}\biggl(\frac{\hbar^2 k^2
}{2m }- \frac{mc^2}{2}\biggl) \biggl (\vec{c}^{+}_{\vec{k},{\bot}}
\vec{c}^{+}_{-\vec{k},{\bot}}+
\vec{c}_{-\vec{k},{\bot}}\vec{c}_{\vec{k},{\bot}}\biggl) +\\[+12pt]
\displaystyle +\; \sum_{0<k^{'}<k_0}\biggl(\frac{\hbar^2 {k^{'}}^2
}{2m }+ \frac{mc^2}{2}\biggl)
\vec{c}^{+}_{\vec{k}^{'},{\bot}}\vec{c}_{\vec{k}^{'},{\bot}}+\\[+12pt]
\displaystyle
+\;\frac{1}{2}\sum_{0<k^{'}<k^{'}_0}\biggl(\frac{\hbar^2 k^2 }{2m }-
\frac{mc^2}{2}\biggl) \biggl (\vec{c}^{+}_{\vec{k}^{'},{\bot}}
\vec{c}^{+}_{-\vec{k}^{'},{\bot}}+
\vec{c}_{-\vec{k}^{'},{\bot}}\vec{c}_{\vec{k}^{'},{\bot}}\biggl)
\end{array}
\end{equation}
where taking into account (134) with introducing new sort
transformations

\begin{equation}
\left.
\begin{array}{c}
r_{\bot}\vec{a}_{\vec{k}} =\frac{r_{\bot}\vec{r}_{\vec{k}} +
M_{\vec{k},{\bot}}r^{\ast}_{\bot}\vec{r}^{+}_{-\vec{k}}} {\sqrt{1-M^2_{\vec{k},{\bot}}}} \\
r_{\mid\mid}\vec{a}_{\vec{k}} =\frac{r_{\mid\mid}\vec{r}_{\vec{k}} +
M_{\vec{k}, {\mid\mid}}r^{\ast}_{\mid\mid}\vec{r}^{+}_{-\vec{k}}}
{\sqrt{1-M^2_{\vec{k},{\mid\mid}}}}\\
\end{array} \right\}
\end{equation}

where $M_{\vec{k},{\bot}}$ and $M_{\vec{k},{\mid\mid}}$ are the real
symmetrical functions of a wave vector $\vec{k}$. Thus, the
Hamiltonian operator of reflected $\hat{H}_{R,r}$ radiation takes
diagonal form:

\begin{equation}
\hat{H}_{R,r}=\hat{H}_{0,r}+\sum_{ 0< k<
k_0}\xi_{\vec{k}}\vec{r}^{+}_{\vec{k}}\vec{r}_{\vec{k}}
\end{equation}
where
\begin{equation}
\hat{H}_{0,r}=mc^2N_0 R(\vec{k}=0,\omega_{\vec{k}=0})
\end{equation}
is the energy of reflected Light-Particles in the condensate;

\begin{equation}
\xi_{\vec{k}}=\hbar k c R(\vec{k},\omega_{\vec{k}})
\end{equation}
is the energy of reflected polaritons; the reflection coefficient is

\begin{equation}
R(\vec{k},\omega_{\vec{k}})=\mid r_{\mid\mid}\mid
^2cos^2\alpha_i+\mid r_{\bot}\mid ^2(\vec{k},\omega_{\vec{k}})
\end{equation}

$\vec{r}^{+}_{\vec{k}}$ and $\vec{r}^{+}_{\vec{k}}$ are,
respectively, the Bose vector-operators of creation and annihilation
of reflected polaritons.

By support of above prescription, it is easy to get the Hamiltonian
operator of refracted $\hat{H}_{R,t}$ radiation which is:

\begin{equation}
\hat{H}_{R,t}=\hat{H}_{0,t}+2\sum_{ 0< k<
k_0}\eta_{\vec{k}}\vec{t}^{+}_{\vec{k}}\vec{t}_{\vec{k}}
\end{equation}

where
\begin{equation}
\hat{H}_{0,t}=mc^2N_0\hat{n}(\vec{k}=0,\omega_{\vec{k}=0})T(\vec{k}=0,\omega_{\vec{k}=0})
\end{equation}

is the energy of refracted Light-Particles in the condensate;

\begin{equation}
\eta_{\vec{k}}=\hbar k c\hat{n}T(\vec{k},\omega_{\vec{k}})
\end{equation}

is the energy of refracted polaritons; the refracted coefficient is
\begin{equation}
T(\vec{k},\omega_{\vec{k}})=\mid t_{\mid\mid}\mid
^2cos^2\alpha_i+\mid t_{\bot}\mid ^2(\vec{k},\omega_{\vec{k}})
\end{equation}

$\vec{t}^{+}_{\vec{k}}$, and $\vec{t}_{\vec{k}}$ are, respectively,
the Bose vector-operators of creation and annihilation of refracted
polaritons.

In these terms, the intensities of incident, reflection and
refraction waves take the forms:

\begin{equation}
I_i=2c \hat{H}_{R,i}cos\theta_i
\end{equation}

\begin{equation}
I_r=2c \hat{H}_{R,r}cos\theta_i
\end{equation}

and

\begin{equation}
I_t=2c \hat{H}_{R,t}cos\theta_t
\end{equation}

Hence we note that from the law conservation energy:
\begin{equation}
R+G=1
\end{equation}

where the ratio $R=\frac{I_r}{I_i}$ is the reflection capability;
the ratio $G=\frac{I_t}{I_i}$ is the transmission capability.

As result our investigation, the energies of incidence
$\hat{H}_{0,i}$ (144), reflected $\hat{H}_{0,r}$ (151) and refracted
$\hat{H}_{o,t}$ (155) Light-Particles in the condensate depend on
the density of Light Particles in the condensate $\frac{N_0}{V}$ in
air. Therefore, it needs to treat the BEC of Light Bosons in vacuum

\vspace{5mm}

{\bf VIII. BEC of Light Bosons in vacuum.}

\vspace{5mm}

The connection between the ideal Bose gas and superfluidity in
helium $^4$He was first made by London [2] in 1938. He postulated
that the ideal Bose gas undergoes a phase transition at sufficiently
low temperatures to a condition in which the zero-momentum quantum
state is occupied by a finite fraction of the atoms $\frac{N_0}{N}$
of liquid $^4$He. This momentum-condensed phase was postulated by
London for presentation of the superfluid component of $^4$He. To be
refused a broken of the Bose-symmetry law for the bosons being into
condensate, we apply the Penrose-Onsager's definition of the Bose
condensation [31]:

\begin{equation}
\lim_{N_0, N\rightarrow\infty}\frac{N_0}{N}=const
\end{equation}

Hence in analogy manner, we postulate that a finite fraction of
$\frac{N_0}{N}$ of the Light Bosons are been in the zero-momentum
quantum states. Therefore, we may separate a number $N_0$  of Light
Particles into condensed state by letting

\begin{equation}
\hat{N}_0+\sum_{0< k< k_0}
\vec{a}^{+}_{\vec{k}}\vec{a}_{\vec{k}}=\hat{N}
\end{equation}

In statistical equilibrium, the equation (163) takes a following
form:
\begin{equation}
N_{0, T} +\sum_{0< k< k_0}
\overline{\vec{a}^{+}_{\vec{k}}\vec{a}_{\vec{k}}} =N
\end{equation}

To find the form
$\overline{\vec{a}^{+}_{\vec{k}}\vec{a}_{\vec{k}}}$, we use of
linear transformation presented in (32):
$$
\overline{\vec{a}^{+}_{\vec{k}}\vec{a}_{\vec{k}}}=
\frac{1+L^2_{\vec{p}}}{1-L^2_{\vec{p}}}\overline{\vec{i}^{+}_{\vec{k}}\vec{i}_{\vec{k}}}+
\frac{L_{\vec{k}}}{1-L^2_{\vec{k}}}\biggl(\overline{\vec{i}^{+}_{\vec{k}}\vec{i}^{+}_{-\vec{k}}}
+ \overline{\vec{i}_{\vec{k}}\vec{i}_{-\vec{k}}}\biggl) +
\frac{L^2_{\vec{k}}}{1-L^2_{\vec{k}}}
$$

where $\overline{\vec{i}^{+}_{\vec{k}}\vec{i}_{\vec{k}}}$ is the
average number of photons  with the wave vector $\vec{k}$ at
temperature $T$, presented by (46).

By the theorem of the Bloch-De-Dominisis, we have

\begin{equation}
\overline{\vec{i}^{+}_{\vec{k}}\vec{i}^{+}_{-\vec{k}}}=
\overline{\vec{i}_{\vec{k}}\vec{i}_{-\vec{k}}}=0
\end{equation}

Consequently, the equation for the density of Light Particles in the condensate
takes a following form:

\begin{equation}
\frac{N_{0, T} }{V}=\frac{N }{V}- \frac{1}{V}\sum_{0< k<
k_0}\frac{L^2_{\vec{k}}}{1-L^2_{\vec{k}}}- \frac{1}{V}\sum_{ 0< k<
k_0}\frac{1+L^2_{\vec{k}}}{1-L^2_{\vec{k}}}
\overline{\vec{i}^{+}_{\vec{k}}\vec{i}_{\vec{k}}}
\end{equation}

where the real symmetrical function $L_{\vec{k}}$ from a wave vector
$\vec{k}$ which equals to

\begin{equation}
L^2_{\vec{k}}=\frac{\frac{\hbar^2 k^2}{2m
}+ \frac{mc^2}{2}-\hbar k c}{\frac{\hbar^2
k^2}{2m }+ \frac{mc^2}{2}+\hbar k
c}
\end{equation}

As we have been seen in above the plasmon-polariton and resonance
effects are fulfilled near the condensate level $\vec{k}=0$. This
fact claims to estimate the density Light bosons $\frac{N_{0}}{V}$
in the condensate at approximation $\vec{k}\rightarrow 0$. This
request leads to vanishing the sums in right side of Eq.(166), and
in turn $\frac{N_{0} }{V}\approx \frac{N }{V}$, at a temperature
$T$.

\vspace{5mm}

{\bf IX. Experimental confirmation of existence Light Particles.}

\vspace{5mm}

Now we consider the following cases:

1. In the case of the plasmon-polariton excitations around the
plasmon frequency, we have a refractive index $\hat{n}(\vec
{k},\omega_{\vec {k},p})=\frac{kc}{\omega_p}$, at
$\omega_{\vec{k}}\sim\omega_p$. From the law of refraction, the
incident angle must be $\theta_i=0$, from (130). In this case, we
may consider $\theta_t=0$ because condition $\theta_t=\frac{\pi}{2}$
is absent due to (130). Thus, at $\theta_i=0$ and $\theta_t=0$ with
approximation $\vec{k}\rightarrow 0$, we substitute $\hat{n}(\vec
{k},\omega_{\vec {k}})=\frac{kc}{\omega_p}$ into (136)-(139). Then,

\begin{equation}
r_{\bot}=\frac{1 -\frac{kc}{\omega_p}}{1
+\frac{kc}{\omega_p}}\approx 1
\end{equation}

\begin{equation}
r_{\mid\mid}=-\frac{1 -\frac{kc}{\omega_p}}{1
+\frac{kc}{\omega_p}}\approx -1
\end{equation}
and
\begin{equation}
t_{\bot}=\frac{2}{1  +\frac{kc}{\omega_p}}\approx 2
\end{equation}

\begin{equation}
t_{\mid\mid}=\frac{2 }{\frac{kc}{\omega_p} + 1}\approx 2
\end{equation}

Obviously, the reflection and refraction coefficients are $R=1$ and
$T=4$, therefore, the Hamiltonian operators of reflected
$\hat{H}_{R,r}$ the plasmon-polariton excitations around the plasmon
frequency $\omega_p$ take the form:

\begin{equation}
\hat{H}_{R,r}=\hat{H}_{0,r}+\sum_{ 0<k<
k_0}\xi_{\vec{k}}\vec{r}^{+}_{\vec{k}}\vec{r}_{\vec{k}}
\end{equation}
where
\begin{equation}
\hat{H}_{0,r}=mc^2N
\end{equation}
is the energy of reflected Light-Particles in the condensate;

\begin{equation}
\xi_{\vec{k}}=\hbar k c
\end{equation}
is the energy of reflected plasmon-polariton excitations.

The Hamiltonian operator of refracted $\hat{H}_{R,t}$
plasmon-polariton excitations, around the plasmon frequency
$\omega_p$, takes the form:

\begin{equation}
\hat{H}_{R,t}=2\sum_{  k<
k_0}\eta_{\vec{k}}\vec{t}^{+}_{\vec{k}}\vec{t}_{\vec{k}}
\end{equation}
where the energy of refracted plasmon-polariton excitations

\begin{equation}
\eta_{\vec{k}}=\frac{\hbar^2k^2}{2m_l}
\end{equation}

with effective mass
$$
m_l=\frac{\hbar\omega_p}{8c^2}\approx 10^{-6}m_e
$$

As we see the incoming incidence light induces the
Bose-quasiparticles with an effective mass $m_l\approx
\cdot10^{-6}m_e$ that is small in regard to one in metal with
effective mass $m_p=\frac{\hbar\omega_p}{2c^2}\approx
5\cdot10^{-6}m_e$ (117).

2. In the case of the resonance-polariton excitations, the
refractive index is $\hat{n}(\vec
{k},\omega_{\vec{k},r})=\frac{2m_0c}{\hbar k}$, around the resonance
frequency $\omega_{\vec {k},r}=\frac{\hbar k^2}{2m_0}$, with the
approximation $\vec{k}\rightarrow 0$. Then, we may consider two
important cases: Case 1. $\theta_i=0$ and $\theta_t=0$; Case 2.
$\theta_i\not=0$, $\theta_i\not=\frac{\pi}{2}$ and
$\theta_t=\frac{\pi}{2}$. Substituting $\hat{n}(\vec
{k},\omega_{\vec {k}})=\frac{2m_0c}{\hbar k}$ into (136)-(139),
gives the result for case 1. $\theta_i=0$ and $\theta_t=0$,

\begin{equation}
r_{\bot}=\frac{1 -\frac{2m_0c}{\hbar k}}{1 +\frac{2m_0c}{\hbar
k}}\approx 1
\end{equation}

\begin{equation}
r_{\mid\mid}=-\frac{\frac{2m_0c}{\hbar k} -1}{\frac{2m_0c}{\hbar k}
 +1}\approx -1
\end{equation}

and

\begin{equation}
t_{\bot}=\frac{2}{1 +\frac{2m_0c}{\hbar k}}\approx \frac{\hbar
k}{m_0c}
\end{equation}

\begin{equation}
t_{\mid\mid}=\frac{2}{\frac{2m_0c }{\hbar k} + 1}\approx \frac{\hbar
k}{m_0c}
\end{equation}
Obviously, the reflection coefficient is $R=1$. Therefore, the
Hamiltonian operators of reflected $\hat{H}_{R,r}$
resonance-polariton excitations, around the resonance frequency
$\omega_{\vec {k},r}$, takes the form:

\begin{equation}
\hat{H}_{R,r}=\hat{H}_{0,r}+\sum_{ 0<k<
k_0}\xi_{\vec{k}}\vec{r}^{+}_{\vec{k}}\vec{r}_{\vec{k}}
\end{equation}
where
\begin{equation}
\hat{H}_{0,r}=mc^2N
\end{equation}
is the energy of reflected Light-Particles in the condensate;

\begin{equation}
\xi_{\vec{k}}=\hbar k c
\end{equation}
is the energy of reflected resonance-polariton excitations.

The Hamiltonian operator of refracted $\hat{H}_{R,t}$
resonance-polariton excitations, around the resonance frequency
$\omega_{\vec {k},r}$, takes the form:

\begin{equation}
\hat{H}_{R,t}=2\sum_{  k<
k_0}\eta_{\vec{k}}\vec{t}^{+}_{\vec{k}}\vec{t}_{\vec{k}}
\end{equation}
where the energy of the refracted resonance-polariton excitations
appear as Bose-quasiparticles

\begin{equation}
\eta_{\vec{k}}=\frac{\hbar^2k^2}{2m_r}
\end{equation}

with effective mass
$$
m_r=\frac{m_0}{4}=0.5\cdot m_e
$$

As we see the case, when $\theta_i=0$ and $\theta_t=0$ cannot lead
to presence of Light Particles in the condensate which is needed to
explain the problem of the enhancement of intensities of Light
experiments SERS and metal films.

Now, we consider an interesting case $\theta_i\not=0$,
$\theta_i\not=\frac{\pi}{2}$, $\theta_t=\frac{\pi}{2}$ and
$\alpha_i=\frac{\pi}{2}$, when the vector of electric field
$\vec{E}_i$ of incident Light is perpendicular to the plane of
incidence. Then, $R=1$ but

\begin{equation}
t_{\bot}\approx 2
\end{equation}

Then, the quadrat of refracted coefficient is  $ T=4$, and in turn, the
Hamiltonian operator of refracted $\hat{H}_{R,t}$
resonance-polariton excitations, around the resonance frequency
$\omega_{\vec {k},r}$, take the form:

\begin{equation}
\hat{H}_{R,t}=\hat{H}_{0,t}+2\sum_{0<k<
k_0}\eta_{\vec{k}}\vec{t}^{+}_{\vec{k}}\vec{t}_{\vec{k}}
\end{equation}
where the energy of refracted resonance-polariton excitations appear
as Bose-quasiparticles:

\begin{equation}
\eta_{\vec{k}}=8m_0 c^2
\end{equation}
and the energy of Light Particles in the condensate into metal
\begin{equation}
\hat{H}_{R,0}=\lim_{k\rightarrow 0}\frac{8m_0 m c^3 N}{\hbar
k}\rightarrow \infty
\end{equation}
This result is a very important because it shows the existence of a
constant electric field near the surface of metal.

As we know, the  energy of Light-Particles of incident Light in the
condensate is $\hat{H}_{0,i}=mc^2N$, which determines the constant
electric field $\vec{E}_i$ of incident light on the surface of
metal, may be define as:

\begin{equation}
\frac{E^2_i}{8\pi}=\frac{mc^2 N }{V}
\end{equation}

In this respect, the energy of transmitted Light Particles in the
condensate, which are concentrated near the interface of surface
metal, represents the energy of constant electric field $\vec{E}_t$
of transmitted light by following formulae:
\begin{equation}
\frac{E^2_t}{8\pi}=\frac{8m_0 m c^3 N}{\hbar k V}
\end{equation}

The ratio
\begin{equation}
\frac{E^2_t}{E^2_i}=\frac{8m_0 c}{\hbar k}
\end{equation}

This result confirms the result of experiments [8-10] where highly
unusual transmission properties have been shown for metal films
perforated with a periodic array of subwavelength holes, because the
electric field is highly localized inside the grooves (around
300-1000 times larger than intensity of incoming optical light). On
other hand, it seems the ratio $\frac{E^2_t}{E^2_i}$ does not depend
on the mass of Light Particles $m$. However, at a maximum quantity
of wave number $k=k_0=\frac{mc}{h}$, presented in (44), and a
minimum quantity of the ratio $\frac{E_t}{E_i}=300$, we may find the
mass $m=\frac{16 m_e}{9\cdot 10^4}=1.8\cdot 10^{-4} m_e$ from
Eq.(192) which confirms the existence of Light Particles. Thus, the
presence of the Light Particles in the condensate provides the
launching of the surface Fr$\ddot o$lich-Schafroth bosons on the
surface metal holes, at a ratio $\frac{E_t}{E_i}=300-1000$.

\vspace{5mm}

{\bf X. Conclusion.}

\vspace{5mm}

In conclusion, we show the advantages  of the resonance-polariton
model in comparison with the plasmon-polariton model. So, 1. the
model plasmon-polariton is analysed at $T=0$ but the model
resonance-polariton at any temperature $T$; 2. the mass of the
plasmon-polariton is $m_p\approx 10^{-6}m_e$ but the mass of the
resonance-polariton $m_r\approx 0.5m_e $, which implies the
resonance-polariton has a heavy mass relative to the mass of Light
Particle. This is the reason to suggest that  the model surface
plasmon-polariton cannot explain the observation results of the
experiments connected with optical transmission through metal films,
where the electric field is highly localized [8-10] as well as SERS
experiments [12,13]. Indeed, the so-called  resonance effect can
explain these experiments; 3. in the case of plasmon-polariton
excitations, the transmitted Light cannot be propagated by direction
along of surface metal, and therefore, there are not the Light
Particles in the condensate into metal which is needed to explain
the problem of the enhancement of optical property of metal surface
but in the case of resonance-polariton excitations, it is a
possible.

In fact, the resonance effect is novel because it does not depend on
temperature $T$. Hence, we note that the resonance effect connected
with the dielectric response of medium considering by Bose gas
consisting of the Fr$\ddot o$lich-Schafroth charged singlet bosons
with charge $e_0=2e$ and mass $m_0=2 m_e$.

\vspace{5mm}

{\bf Acknowledgements.}

\vspace{5mm}

We are particularly grateful to Professor Marshall Stoneham F R S
(London Centre for Nanotechnology, and Department of Physics and
Astronomy University College London, Gower Street, London WC1E 6BT,
UK) for help with the English.

\newpage
\begin{center}
{\bf References}
\end{center}

\begin{enumerate}

\item
Minasyan V. N. and Samoilov V. N. Two  Type Surface Polaritons
Excited into Nanoholes in Metal Films, Progress in Physics, {\bf 2},
3-6 (2010)
\item
F.London, Nature, {\bf 141}, 643 (1938)
\item
L.Landau, Journal of Physics (USSR) {\bf 5}, 77 (1941); Journal of
Physics (USSR) {\bf 11}, 91 (1947).
\item
N.N.Bogoliubov, Journal of Physics (USSR) {\bf 11}, 23-32 (1947).
\item
Dirac~P.A.M. The Principles of Quantum Mechanics. Clarendon press,
Oxford, (1958).
\item
de~Broglie~L. Researches on the quantum theory. Annalen der Physik,
{\bf 3}, 22--32 (1925)
\item
A.H.~Compton~, Phys.Rev., {\bf 21}, 483 (1923).
\item
Lopez-Rios ~T.,Mendoza~D.,Garcia-Vidal~F.J.,
Sanchez-Dehesa~J.,Panneter~B. Physical Review Letters, {\bf 81} (3),
665-668 (1998).
\item
Ghaemi ~H.F.,Grupp~D. E.,Ebbesen~T.W., Lezes~H.J. Physical Review B,
{\bf 58} (11), 6779-6782 (1998).
\item
Sonnichen  C.,Duch A.C,Steininger G., Koch~M.,Feldman~J. Applied
Physics Letters,  {\bf 76},(2), 140-142 (2000).
\item
Raether H. Surface plasmons. Springer-Verlag, Berlin, (1988).
\item
Nie, S.; Emory, S. R., Probing Single Molecules and Single
Nanoparticles by Surface-Enhanced Raman Scattering., Science, 275,
(5303), 1102-1106 (1997).
\item
Fleischmann, M.; PJ Hendra and A.J McQuillan., "Raman Spectra of
Pyridine Adsorbed at a Silver Electrode". Chemical Physics Letters
26 (2), 163–166 (1974)
\item
Born~M. and Wolf~E. Principles of Optics. Pergamon press, Oxford,
(1980).
\item
Huang K., "Statistical Mechanics", John Willey, New York-London,
(1963).
\item
Minasyan V,N.  Light Bosons of Electromagnetic Field  and Breakdown
of Relativistic Theory. arXiv::0808.0567 (2009).
\item
Kubo R., Lectures in Theoretical Physics, Vol.1 (Boulder) (Wiley-
Interscience, New York, (1959), 120-203. Gerald D.~Mahan~,
"Many-Particle Physics", ~Plenium~ Press New York, (1990).
\item
Minasyan V. N. and Samoilov V. N., Physics Letters A 374, 2792
(2010).
\item
Minasyan V. N. and Samoilov V. N. Two  Formation of Singlet Fermion
Pairs in the Dilute Gas of Boson-Fermion mixture, Progress in
Physics., {\bf 4}, 3-9 (2010).
\item
H. Fr$\ddot o$lich, Proc.Roy. Soc, A215, 291 (1952).
\item
E.~Maxwell~, Phys.Rev., {\bf 78}, 477 (1950); {\bf 79}, 173 (1950).
\item
Meissner, W.; R. Ochsenfeld  "Ein neuer effekt bei eintritt der
supraleitfahigkeit", Naturwissenschaften {\bf 21} (44): 787–788
(1933).
\item
M.R.~Schafroth~, Phys.Rev., {\bf 100}, 463 (1955).
\item
L.N.~Cooper~, Phys.Rev., {\bf 104}, 1189 (1956).
\item
J.~Bardeen~, L.N.~Cooper~, and J.R.~Schrieffer~, Phys.Rev.{\bf
108},~1175~(1957); N.N.~Bogoliubov~, ~Nuovo~ ~Cimento~, ~{\bf
7},~794~(1958)
\item
L. L. Foldy, Physical Review, {\bf 124}, 649 (1961).
\item
R. F. Bishop, J. Low. Temp. Phys., {\bf 15}, 601 (1974).
\item
S.R.Hore, N.E.Frenkel, Physical Review B {\bf 12}, 2619 (1975).
\item
F. Abeles "Optical property of solids", Publishing company
Amsterdam-London, (1972).
\item
D. Pines., "The Many-Body Problem", (W. A. Benjamin: N.Y) (1961).
\item
Penrose O., Onsager  L.,  Physical Review  {\bf 104}, 576  (1956).

\end{enumerate}
\end{document}